\documentclass[12pt]{iopart}
\usepackage[utf8]{inputenc}
\usepackage[T1]{fontenc}

\usepackage{hyperref}
\hypersetup{
    colorlinks=true,
    linkcolor=blue,
    filecolor=magenta,      
    urlcolor=cyan,
    citecolor=blue}

\usepackage{graphicx}
\usepackage[justification=centering]{caption}
\usepackage[justification=centering]{subcaption}

\usepackage{siunitx}
\DeclareSIUnit\gauss{G}

\usepackage{booktabs}
\newcolumntype{C}{>{$}c<{$}}
\newcolumntype{L}{>{$}l<{$}}
\newcolumntype{R}{>{$}r<{$}}

\usepackage[noabbrev]{cleveref}
\usepackage[dvipsnames]{xcolor}
\usepackage{todonotes}

\begin{document}


\title[Electron heating and energization in rfMS studied via a 2d3v PIC/MCC code]{Electron dynamics in planar radio frequency magnetron plasmas: II. Heating and energization mechanisms studied via a 2d3v particle-in-cell/Monte Carlo code}


\author{D. Eremin$^{1}$, B. Berger$^{2}$, D. Engel$^{1}$, J. Kallähn$^{1}$, K. Köhn$^{1}$, D. Krüger$^{1}$, L. Xu$^{1}$, M. Oberberg$^2$, C. Wölfel$^3$, J. Lunze$^3$, P. Awakowicz$^2$, J. Schulze$^{2}$, R.P. Brinkmann$^{1}$}

\address{$^1$ Institute of Theoretical Electrical Engineering, Ruhr University Bochum, Universitätsstrasse 150, D-44801 Bochum, Germany}
\address{$2$ Institute of Applied Electrodynamics and Plasma Technology, Ruhr University Bochum, Universitätsstrasse 150, D-44801 Bochum, Germany}
\address{$3$ Institute of Automation and Computer Control, Ruhr University Bochum, Universitätsstrasse 150, D-44801 Bochum, Germany}

\ead{denis.eremin@rub.de}
\vspace{10pt}
\begin{indented}
\item[]
\today
\end{indented}

\begin{abstract}
The present work investigates electron transport and heating mechanisms using an $(r,z)$ particle-in-cell (PIC) simulation of a typical rf-driven axisymmetric magnetron discharge with a conducting target. Due to a strong geometric asymmetry and a blocking capacitor, the discharge features a large negative self-bias conducive to sputtering applications. Employing decomposition of the electron transport parallel and perpendicular to the magnetic field lines, it is shown that for the considered magnetic field topology the electron current flows through different channels in the $(r,z)$ plane: a ``transverse'' one, which involves current flow through the electrons' magnetic confinement region (EMCR) above the racetrack, and two ``longitudinal'' ones, where 
electrons' guiding centers move along the magnetic field lines. Electrons gain energy from the electric field along these channels following various mechanisms, which are rather distinct from those sustaining dc-powered magnetrons. The longitudinal power absorption involves mirror-effect heating (MEH), nonlinear electron resonance heating (NERH), magnetized bounce heating (MBH), and the heating by the ambipolar field at the sheath-presheath interface. The MEH and MBH represent two new mechanisms missing from the previous literature. The MEH is caused by a reversed electric field needed to overcome the mirror force generated in a nonuniform magnetic field to ensure sufficient flux of electrons to the powered electrode, and the MBH is related to a possibility for an electron to undergo multiple reflections from the expanding sheath in the longitudinal channels connected by the arc-like magnetic field. The electron heating in the transverse channel is caused mostly by the essentially collisionless Hall heating in the EMCR above the racetrack, generating a strong ${\bf E \times B}$ azimuthal drift velocity. The latter mechanism results in an efficient electron energization, i.e., energy transfer from the electric field to electrons in the inelastic range. Since the main electron population energized by this mechanism remains confined within the discharge for a long time, its contribution to the ionization processes is dominant.
\end{abstract}

%
\vspace{2pc}
\noindent{\it Keywords}: magnetized plasma, magnetized rf discharge, magnetron, electron heating, electron transport

\pacs{52.25.-b,52.25.Xz,52.27.-h,52.27.Aj,52.50.Dg,52.50.Qt,52.55.-s,52.77.-j,52.80.Pi} 


%


%
%
%


\section{Introduction} \label{sec1}

The physical vapor deposition (PVD) based on either dc- or rf-driven magnetron sputtering (dcMS and rfMS) \cite{lieberman_2005,thornton_1978,waits_1978,thornton_1981}
is a well-established technique
that uses a magnetic field and has a long 
record of successful applications in the plasma processing technologies. 
The dcMS and rfMS techniques are similar in terms of the technical realization and can be utilized on the same device, which is beneficial when deposition of layered
films is considered (e.g., \cite{gulkowski_2020}). However, they are quite different in terms of the physics and applications. The rfMS has its own
advantages and disadvantages. Among the former one can count the ability to use it for both conducting and insulating targets, and
the stability of operation with any type of process gas. 
A sinusoidal driving voltage  helps to avoid the charge accumulation at reactor surfaces causing 
microarc discharges \cite{minea_2003b}. 
Because of this, rfMS can sustain a larger dc sheath voltage fall and lead to higher energies of ions impinging on the substrate compared to dcMS, resulting in reduced stress and better conductivity of the deposited film \cite{ellmer_1998}. 
One of the major diasadvantages of rfMS is a lower deposition rate \cite{ellmer_1998,alfonso_2012,gudmundsson_2020}
compared to dcMS.
Note that dcMS and rfMS can also be employed simultaneously, where rfMS is enhanced by supplying an external dc voltage source, rather than using the self-bias
produced at a blocking capacitor. In this way, one can combine merits of both technologies \cite{bender_1999,stowell_2007,nomoto_2011,liu_2016}.

The mentioned magnetized discharges are typically operated at a low neutral gas pressure to minimize the interaction of heavy particles produced in
the discharge 
with neutral particles of the working gas and its products, which would lead to undesired energy losses and broadening of the ion velocity angle distribution.
Such discharges thus have to rely on the magnetic field to ensure sufficient power absorption for their sustainment in spite of the low neutral gas pressure. The magnetic field utilized for this purpose typically has a strength sufficient only for the electron magnetization whereas the magnetic field's influence on the ions can be neglected.
While the magnetic field confines electrons by inhibiting their motion across the magnetic field lines, electrons are essentially free to move
along the magnetic field lines until they are reflected either by the sheath electric field or by the mirror force
caused by the magnetic moment's conservation (e.g., \cite{lieberman_2005}). In general, any discharge needs an electric field driving a current
powering the discharge. In combination with the magnetic field, the electric field commonly results in a strong ${\bf E}\times{\bf B}$ drift for electrons. The reactor geometry is typically constructed in such a way that the resulting electron drift orbits are closed and do not cross any reactor surface \cite{thornton_1978,lin_1984}.
  
The production of new particles is vital for the functioning of any plasma discharge, since it compensates particle losses due to recombination
processes in the plasma bulk and at the reactor walls. It is one of the key factors determining the plasma density (and hence the ion flux),
both in terms of the average value and the spatial profile.  
It is, therefore, important to understand how electrons acquire energy from
the electric field and how that energy is distributed between them \cite{zheng_2021}, producing low- and high-energy electrons. The electron power absorption
is intimately related to the electron transport along the electric field due to the fact that the absorbed power density is equal to $({\bf j_e}\cdot{\bf E})$, where ${\bf j_e}$ is
the electron current density. The electron transport can occur either along the magnetic field or orthogonal to it, and in the latter case the electron mobility is strongly influenced by
the magnetic field. The electric field powering the discharge is often generated perpendicular to the magnetic field, so that it is the electron transport orthogonal to the magnetic field lines which should produce the finite ${\bf j_e}$ component along the electric field. 
Due to the enhanced power absorption and ionization efficiency in magnetized capacitively coupled radio-frequency (CCRF) discharges compared to their DC counterparts, 
the magnetic fields that are normally used in rfMS 
are relatively weak (< 20 mT). Therefore, here we will assume that the electron transport perpendicular to the magnetic field lines
is dominated by classical mechanisms - the collision and polarization (inertial) drifts \cite{lieberman_2005}
, since instabilities causing the anomalous transport typically emerge for larger magnetic fields
\cite{panjan_2019,lucken_2020,xu_2021,xu_2022}. The anticipated absence of such instabilities allows us to adopt the azimuthal symmetry and resolve only the radial and axial
coordinates. By doing this, we assume that breaking of the azimuthal symmetry by experimental equipment, for example, due to an azimuthal magnetic field modulation, can also be neglected.

The difference in the deposition rates in dcMS and rfMS for a fixed power is directly related to their power absorption or ionization efficiency \cite{ellmer_1998}, which can be
interpreted as the efficiency of channelling the power absorbed by electrons to their high-energy fraction capable of participation in ionization events. 
Following \cite{anders_2014}, we will also employ the term ``electron energization'' to refer to the process of energy transfer from the electric field to electrons in the inelastic energy range
(as opposed to the more general ``electron heating'' associated with the net increase of kinetic energy for all electrons). The term ``inelastic energy range'' is understood here as the range of high energies comparable to or greater than the ionization energy of the working gas.  
Focusing on weak magnetic fields and, therefore, dropping the electron heating mechanisms caused by instabilities and the associated self-organized structures such as spokes \cite{anders_2014},
the electron power absorption in dcMS 
is mainly related to the acceleration of secondary electrons in the cathode fall \cite{thornton_1978} and the presheath heating claimed to be of Ohmic nature \cite{huo_2013}.
Hence, dcMS discharges usually require relatively large voltages to be sustained. By the analogy with unmagnetized glow discharges, one can assume that the heating of rf discharges is much more versatile than that in the dc counterparts
and leads to more efficient power absorption. Therefore, for a fixed power, an rfMS discharge would exhibit a significantly smaller voltage amplitude (and thus the time-averaged powered sheath voltage)
compared to a dcMS discharge \cite{ellmer_1998}. 
Since the sputtering
applications require low pressures, the average ion energy at the powered electrode is approximately equal to the powered sheath voltage drop \cite{kuypers_1990,jouan_1994}. 
The smaller dc voltage drop in the powered sheath of an rfMS discharge compared to its dcMS analogue results in the above mentioned lower deposition rates observed in rfMS, since the sputtered yield is typically a monotonously growing function of the impinging ions' energy.

Compared to dcMS, the power absorption mechanisms for rfMS discharges are less studied.
The seminal work of Lieberman et al. \cite{lieberman_1991} (see also \cite{lieberman_2005}) considered a 1D analytical model of a magnetized CCRF discharge and derived analytical expressions
for the absorbed power due to the Ohmic and the sheath heating mechanisms modified by the presence of a magnetic field. In the latter case
it was argued that the sheath heating is significantly enhanced because the cyclotron rotation due to the magnetic field returns the electrons
to collide with the expanding sheath, so that this energy-increasing interaction occurs many times during the sheath expansion. Although the corresponding mechanism was still called ``stochastic heating'' in \cite{lieberman_1991}, we will refer to it as ``magnetized stochastic heating'' (MSH) to differentiate it from the stochastic heating in unmagnetized plasmas. Another electron heating mechanism related to the magnetic field and based on the resonance between the sheath motion and the Larmor rotation leading to the condition $\omega = 2\,\omega_{\rm ce}$ was investigated in \cite{OkunoOhtsuFujita1994} and named there ``electron sheath resonance''. Recently, it has been revisited in a number of papers \cite{Zhang2021,Patil2022,Sharma2022}, where it acquired a new name ``electron bounce-cyclotron resonance heating''.
This effect, however, requires rather weak magnetic fields (e.g., for the driving frequency of $13.56$ MHz studied in the present work the resonance condition leads to $B=0.24$ mT). Therefore, this mechanism will not be considered here.
In later works \cite{hutchinson_1995,turner_1996}, based on results of numerical simulations it was claimed that the collisional heating
ascribed to the Ohmic nature is also amplified by the magnetic field and should actually prevail over all other electron heating mechanisms
for a sufficiently large magnetic field. A more recent work \cite{zheng_2019} also attributed the increase of the power absorption rate with growing magnetic field to the Ohmic heating mechanism and emphasized the fact that it was predominantly due to the Hall current. The analysis of electron power absorption in rfMS discharges was conducted in works \cite{minea_2003b} and \cite{bretagne_2003} as well.
The former considered the rf period-averaged profile of the power density absorbed by electrons obtained from numerical simulations of rfMS in planar magnetron geometry and identified three peaks -
two of them above the race track, which were attributed to the secondary electrons and the Ohmic heating by the azimuthal current, and one peak close to the cathode
surface where the magnetic field was almost perpendicular to the target, which was ascribed to the ``wave-riding'' mechanism related to the sheath motion. 
In the latter work it was shown that  
experimental data demonstrated enhanced ionization efficiency at low pressures, which indicated existence of some heating mechanism of a different nature from the
Ohmic heating, the latter expected to decrease in weakly collisional plasmas. 

This work is a part of a companion paper series (see also \cite{eremin_2021,berger_2021}) aimed at deepening the understanding of electron heating/energization mechanisms in electropositive rfMS discharges with a conducting target.

First, in \cite{eremin_2021} we have used a reduced 1D geometry relevant to a cylindrical magnetron to study mechanisms which could be responsible for the ionization efficiency amplification at lower pressures. 
It was argued that 
to overcome the magnetic field's confining action on electrons and to ensure a sufficient level of their transport
needed to sustain the discharge, the discharge builds up an enhanced electric field, which in turn leads to efficient electron energization. 
Conversely, the electrons close to the powered electrode are inhibited by the magnetic field to move in the radial direction and
can experience a strong electric field at the expanding sheath edge or
a strong reversed electric field \cite{kushner_2003,wang_2020} during the sheath collapse. In addition to the electric field time modulation on the scale of the rf-period, there is an additional time modulation of the electric field on the cyclotron scale, which is due to the change of the electric field over the Larmor radius and thus is a finite-Larmor-radius (FLR) effect.
Similar to previous works \cite{lieberman_1991,hutchinson_1995,turner_1996,zheng_2019} we have concluded that the related dominant electron heating mechanism was due to the Hall current. However,
the corresponding mechanism was argued to be of a conceptually different nature
from the conventional Ohmic heating, which resorts to collisional diffusion in the energy space to produce energetic electrons. In contrast to that, in the Hall heating at low pressure the electron energization occurs via essentially collisionless generation of a strong time-dependent ${\bf E}\times{\bf B}$ azimuthal drift through the aforementioned strong time-dependent electric fields, shifting a large number of electrons into the inelastic energy range (see \cite{eremin_2021} for more details). 
Therefore, such a mechanism should cause an efficient ionization at low pressures.

In the present work we simulated the discharge behaviour for a single set of typical parameters in a more complicated 2D $(r,z)$ geometry of a planar magnetron with a realistic magnetic field topology. Many aspects remain similar to \cite{eremin_2021}. In particular, we find that the dominant heating mechanism is still caused by the Hall heating in the EMCR above the racetrack, where electrons are confined by the magnetic field across, and by the electrostatic sheath potential or by the mirror effect along the magnetic field lines. During the sheath collapse or during the sheath expansion strong electric fields needed to increase the electron transport through this region build up and lead to the Hall heating. However, there are new insights related to the electron motion parallel to the magnetic field lines, which is possible in the considered geometry. A new mechanism responsible for the electric field reversal during the sheath collapse is argued to be caused by the mirror force acting along the magnetic field lines on electrons during the sheath collapse (at the end of the arcing magnetic field lines, where they are almost orthogonal to the target). However, because the electric field is now directed parallel to the magnetic field lines, it does not
contribute to the Hall heating mechanism. Rather, the corresponding electric field increases the longitudinal velocity to let a sufficient number of electrons reach the powered electrode to compensate the positive ion flux on time average (we propose to call such a mechanism ``mirror-effect heating'', or MEH). Further, there is a strong electron heating observed at the moment when the powered sheath starts to expand along the magnetic field lines. This heating can be linked to the nonlinear electron resonance heating (NERH) and the related plasma series resonance (PSR) \cite{mussenbrock_2006,mussenbrock_2008,lieberman_2008,oberberg_2019} mechanism and also increases mostly the longitudinal velocity, so that the majority of the accelerated electrons leave the discharge. Some of them are sufficiently deflected and trapped by the magnetic field, so that they can follow the arc-like magnetic field and collide with the expanding sheath on the other side of the EMCR region, leading to an enhancement of the sheath energization efficiency. We suggest to call the corresponding mechanism enabled by the magnetic field ``magnetized bounce heating" or MBH. 
The electrons captured by the magnetic field replenish the EMCR electron population.
Similar to \cite{zheng_2021}, we observe that the heating due to secondary electrons seems to play only an auxiliary role.

The third companion paper \cite{berger_2021} completes the series by reporting the experimental observations and compares them to the results of 2d3v PIC/MCC simulations for a number of different cases featuring a magnetic flux density variation and a pressure variation in a planar rf-driven magnetron. The experimental results were obtained 
for the spatio-temporally resolved excitation patterns in various areas of a planar rfMS discharge in argon utilizing the phase-resolved optical emission spectroscopy (PROES) technique. 
The data demonstrated a significantly different behavior of the excitation dynamics in the region close to the racetrack, where the magnetic field lines are essentially parallel to the target, and in the regions close to the powered electrode's radial center and its periphery, where magnetic field lines cross the target almost at a right angle. In the first case the excitation pattern peaked at an axial location away from the target and showed almost no high-frequency oscillations associated with the PSR-like phenomena provided the powered sheath width was small enough (due to the magnetic field or pressure increase) to let a sufficient number of electrons enter the region of a significant magnetization. In the second case the excitation pattern exhibited the generation of beam-like energetic electron beams reminiscent of the unmagnetized CCRF discharges.
These patterns can be explained using results of the first and the present papers.


\section{Numerical model} \label{sec2}

To model the planar rfMS discharge in question we used the ECCOPIC2S-M code tailored to simulations of magnetized plasmas. It is 
a two-dimensional electrostatic member of the in-house developed ECCOPIC code family based on the fully implicit energy-conserving particle-in-cell (ECPIC) method. The ECPIC was originally introduced by \cite{chen_2011,markidis_2011} and adapted to bounded plasmas in \cite{eremin_2021,mattei_2017}. As has been recently shown in \cite{Barnes2021}, this algorithm is robust against excitation of the finite grid instability as long as the electron mean drift velocities are smaller than velocities of their thermal motion. Since in the considered geometry the only strong electron drift occurs in the azimuthal direction whereas the azimuthal electric field is not taken into account, and the electron drift velocities in the $(r,z)$ dimensions are much smaller than their thermal velocities,
the finite grid instability is not an issue in the present work.

Unlike the conventional explicit PIC schemes, an implicit code ensures a self-consistent integration of the fields and particles using the field grid with a practically arbitrary cell size. The cell size can easily exceed the local Debye length, which would normally lead to an excessive numerical heating due to the finite grid instability \cite{hockney_1988,birdsall_2005} in case of an explicit scheme. Additionally, an implicit scheme typically makes it possible to evolve the numerical model in time using a time step larger than the local plasma period. However, this is rarely needed in simulations of technological plasmas, since phenomena occurring on the time scale of the plasma period are typically important. Frequently, the time step size is set by another, smaller time scale where other important phenomena could occur, e.g., cyclotron period. The energy-conserving implicit PIC approach is a special kind of an implicit scheme, which aims at an exact (or, more precisely formulated, to a prescribed accuracy) conservation of the discrete analogue of the total energy in the system for the finite time step used in the model's numerical integration 
\cite{chen_2011,markidis_2011}. If the system is not isolated and exchanges energy with the surrounding environment, the energy conservation should be understood in the sense of the Poynting theorem \cite{brackbill_2015,chacon_2019} or its analogue in the electrostatic approximation of Maxwell's equations \cite{decyk_1982,Eremin2022}. The scheme was augmented with a charge-conservation algorithm accompanied by subcycling in the orbit integration, which simultaneously improves conservation of the linear momentum \cite{chen_2011}. With multi-dimensional PIC codes being quite expensive computationally (and implicit multi-dimensional PIC codes even more so), the code was parallelized on GPU 
using algorithms utilized previously in \cite{mertmann_2011} and \cite{schuengel_2012}. The ECCOPIC2S-M code was benchmarked against other codes in the axial-azimuthal \cite{charoy_2019} and the radial-azimuthal \cite{villafana_2021} 
geometries (note that in the latter publication one of the authors (D.E.) used results of an explicit PIC code. Those results were demonstrated to be close to data produced by the implicit ECCOPIC2S-M code under similar conditions in an unpublished work).

\begin{figure}[h]
\centering
\includegraphics[width=10cm]{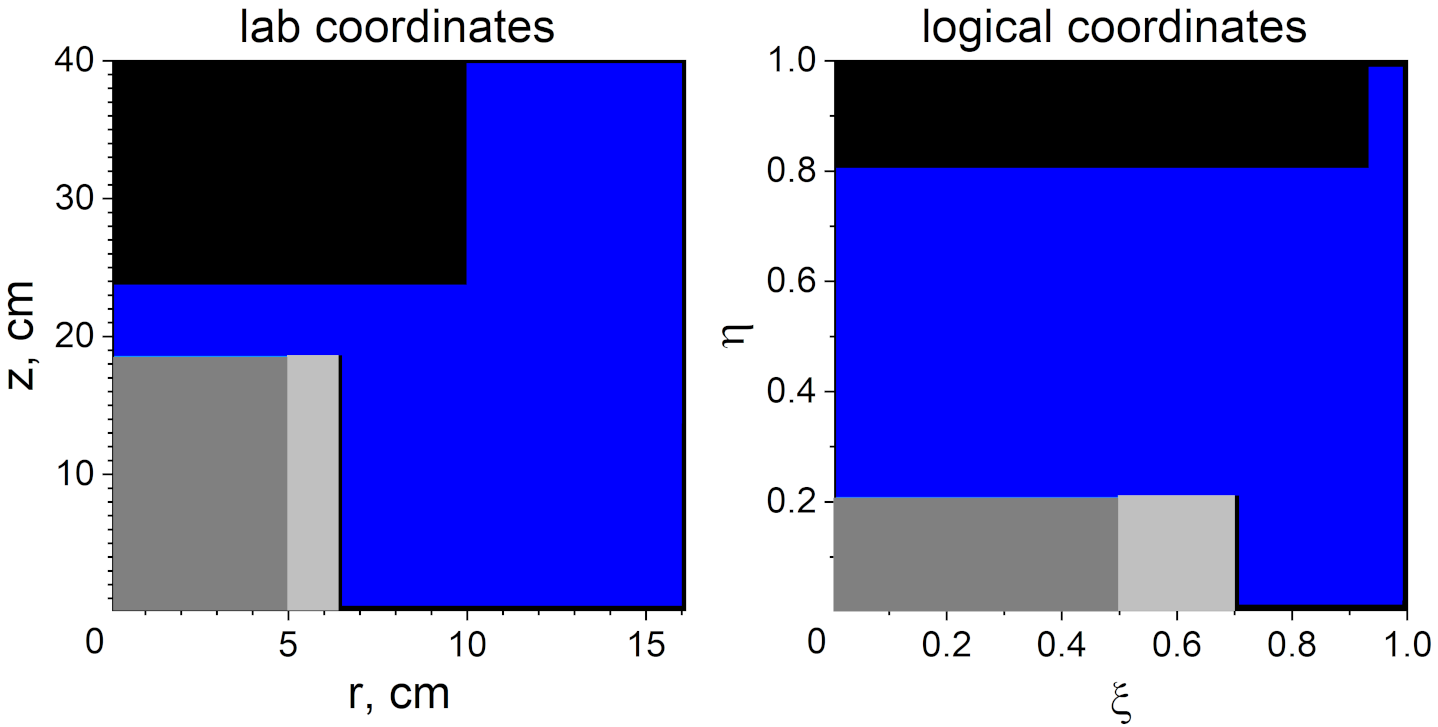}
\caption{Planar magnetron reactor in lab coordinates (left), and in normalized logical coordinates (right). The colors stand for the following parts: blue for the reactor chamber, dark grey for the powered electrode, black for the grounded metal, and light grey for the dielectric spacer extending from $r_{p1}=5$ and $r_{p2}=6.5$ cm and separating the powered and the
grounded parts.}
\label{Fig1}
\end{figure}

A relevant operation regime of rfMS with a conducting target is related to the sputtering applications, which require a large DC self-bias. To produce it, one needs a simple external network with a blocking capacitor sustaining the self-bias, and an asymmetry in the sheath behavior. The reactor geometry
was chosen to model the experimental facility described in \cite{oberberg_2019,oberberg_2018,oberberg_2020}, see Fig.~\ref{Fig1}. 
The reactor chamber can be logically divided into two different parts, the ``main chamber''   
(blue region extending in the radial direction from $r=0$ to $r_{p2}$) and the ''side chamber'' (blue region with $r>r_{p2}$). The discharge plasma is concentrated predominantly in the main chamber, whereas plasma in the side chamber has substantially smaller density.
The asymmetry in the considered setup is dominated by the geometrical factors, since the grounded side chamber is much bigger than the main chamber (see Fig.\ref{Fig1}, left). In the axial direction the main chamber occupies the domain $z_0 = 18.6\,{\rm cm}<z<23.8\,{\rm cm}$.
Even though the side chamber in this reactor is not expected to produce any significant additional plasma, it collects some of the currents produced by the plasma flowing from the main chamber. Therefore, it is important for a proper account of the capacitive voltage division leading to the sheath asymmetry \cite{raizer_1995}. On the other hand, a detailed description of the plasma dynamics in the side chamber in this case is not necessary. If a uniform grid was used for this problem, by far the most part of it would be occupied by the side chamber. The main chamber, which determines the essential plasma dynamics in the system, would make just a tiny fraction of it, resulting in utter computational inefficiency of the algorithm.

This problem can be solved by using nonuniform grids to resolve the areas of interest, while reducing the number of grid cells in less critical parts of the reactor. This can be accomplished by utilizing mapped nonuniform grids without compromising the energy conserving properties of the scheme \cite{chacon_2013}. 
The resulting code can give access to a kinetic and non-local modeling of systems which used to be hard to simulate using conventional PIC schemes \cite{hockney_1988,birdsall_2005}. The strictly enforced conservation of energy is particularly important when utilizing nonuniform grids, since in this case the cell size in the poorly resolved areas can well exceed the Debye length and thus lead to excessive numerical heating if an explicit scheme is employed. Fig.\ref{Fig1}(right) demonstrates how the chosen reactor geometry looks in normalized logical coordinates $(\xi,\eta)$ transformed from cylindrical coordinates $(r,z)$ (with with the corresponding transformation Jacobians $J_r = {\rm d}r/{\rm d}\xi$ and $J_z = {\rm d}z/{\rm d}\eta$). One can clearly see that after such a transformation the main chamber is well resolved whereas the side chamber occupies just a minor part of the grid.
In addition to this, the radial coordinate transformation strongly mitigated the well-known computational difficulty at the rotation axis, which is related to the particle weighting scheme and the electric field solution algorithm in cylindrical coordinates. If one initially loads approximately the same number of superparticles per cell, then the superparticle weight has to be $\propto r J_r$. If one adopts a grid uniform in $\xi = r^2$, then $rJ_r$ is constant and therefore, if no coordinate transformation in the axial coordinate is performed, all superparticles have the same weight. The drawback of this approach is the fact that the grid will be very densely packed close to the radial edge where often no interesting phenomena are expected, resulting in an inefficient grid utilization. A popular alternative is to adopt a grid uniform in $\xi=r$, but in this case superparticle weights are very small close to the radial center and are very large close to the radial edge. In a relatively short time interval the lightweight particles at the axis are replaced by just a few heavyweight particles from the periphery, resulting in a high level of numerical noise at the axis. We propose to use a combination of these two approaches benefiting from the merits of each of them. Using a nonlinear variable transformation, we choose $\xi\propto r^2$ for a few points close to the axis and $\xi\propto r$ for the rest of the radial domain. In this case the superparticle weight remains finite at the axis, whereas the radial grid is efficiently used.
Due to the fact that for the chosen variable transformation $rJ_r$ remains constant at the radial center, the electric field solution algorithm does not have difficulties at the radial center as well.

Due to the strong grid nonuniformity the problem of having superparticles with quite disparate weights remains. The lightweight particles from the center tend to replace the heavyweight particles at the periphery, and their number must grow in order to reproduce even a moderate density there. At the same time, having a small number of heavyweight particles at the center could generate enhanced numerical noise there. Similar problems are caused by the grid non-uniformities in the axial direction, since the grid cell sizes away from the main chamber should be large. These difficulties can be fixed via adaptive particle management techniques consisting in splitting heavyweight superparticles into several lightweight ones and the reciprocal process of merging several lightweight superparticles into fewer heavyweight ones \cite{lapenta_2002,welch_2007,teunissen_2014}. While in collisional plasmas the splitting can be implemented exactly without distorting the velocity distribution functions and satisfying all appropriate conservation laws, the merging algorithm is much more complicated. It must guarantee that by reducing the amount of information about the plasma, essential details of its physics are not eliminated. To this end, a typical merging algorithm chooses particles located close to each other in phase space and attempts to meet several conservation laws (momentum, energy, and charge distribution on the computational grid) so that the merging algorithm's impact on the model remains small. In this work we used a ($4\to 2$) merging algorithm, and we additionally minimized its influence by shifting the merging processes to the side chamber, where no sensitive plasma dynamics was expected to occur, but where the problem of having lightweight particles was particularly severe. A detailed description of the merging algorithm for the ECPIC goes beyond the scope of the present work and will be described elsewhere.

%
%
 
The equations of particle motion for species $s$ in the selected geometry are (e.g., \cite{delzanno_2013b})
\begin{equation}
\left\{
\begin{array}{rll}
m_s\frac{dv_r}{dt} &=& q_s(E_r + v_\theta B_z) + m_s\frac{v_\theta^2}{r} \\
m_s\frac{dv_\theta}{dt} &=& q_s (v_zB_r -v_r B_z) - m_s\frac{v_r v_\theta}{r} \\
m_s\frac{dv_z}{dt} &=& q_s(E_z - v_\theta B_r) \\
\frac{dr}{dt} &=& \frac{v_r}{J_r} \\
\frac{dz}{dt} &=& \frac{v_z}{J_z} . 
\end{array}
\right.   \label{eq2_1}
\end{equation}
Comparing these equations to the ones given in \cite{eremin_2021}, one can see that now equations of motion additionally govern the $z$ coordinate in the configuration and velocity spaces and include the $B_z$ component of the magnetic field. The magnetic field does not influence the particle motion along the magnetic field lines,
whereas it inhibits the transport across them leading to the effects described in \cite{eremin_2021}. To facilitate the charge conservation algorithm of \cite{chen_2011} requiring to stop particles at the logical cell boundaries
we have implemented Eqs.(\ref{eq2_1}) as they are, without resorting to the Boris algorithm typically used for the orbit integration in cylindrical coordinates. As a test, we implemented the Boris algorithm as well, but no significant deviations were detected between the corresponding results obtained in this work, and in the benchmarks mentioned \cite{charoy_2019,villafana_2021}. All superparticles reaching bounding surfaces of the reactor are assumed either to be fully absorbed in case of a metal surface or stuck to it if it was a dielectric. 

In the electrostatic approximation the electric field was derived from the electrostatic potential, and the latter was calculated from Poisson's equation, 
\begin{equation}
\frac{1}{rJ_r}\frac{\partial}{\partial \xi}\frac{\epsilon r}{J_r}\frac{\partial}{\partial \xi} \phi  + \frac{1}{J_z}\frac{\partial}{\partial \eta}\frac{\epsilon}{J_z}\frac{\partial}{\partial \eta}\phi = -e(n_i-n_e),
\label{eq2_2}
\end{equation}
where $\epsilon$ is the material dielectric permittivity (needed to account for the dielectric spacer and the metal walls).  
Then, $E_\xi = -\partial \phi/\partial \xi$ and $E_\eta = -\partial \phi/\partial \eta$. The use of Poisson's equation rather than of the Ampere law's divergence for the electric field update is possible due to the fact that if the charge continuity is strictly enforced (which is the case in ECCOPIC2S-M), then the energy conservation is not violated \cite{chen_2011}. The field components were spatially discretized after Yee's scheme \cite{yee_1966}.

Since the reactor has a nontrivial geometry, to model conducting electrodes in Eq.(\ref{eq2_2}) we set $\epsilon/\epsilon_0$ to a large number, whereas realistic values are used for all other materials. The boundary conditions for the potential $\phi$ are imposed on all boundaries of the rectangular computational domain except the radial center, where the natural boundary condition $E_r=0$ is used. To specify the boundary conditions, a usual assumption that $\phi$ is set to an appropriate potential at each surface (powered electrode was set to $\phi_0$ and all other walls except the dielectric spacer were grounded), is used. At the bottom of the dielectric spacer ($z=0$) a fixed electrostatic potential $\{\phi(r) = \phi_0 \ln(r/r_{p2})/\ln(r_{p1}/r_{p2})\,,\, r_{p1} \leq r \leq r_{p2}\}$ is assumed (Dirichlet boundary condition). 
Such a boundary condition is based on 
an assumption that away from the plasma-dielectric boundary at $z=z_0$ and $r_{p1} \leq r \leq r_{p2}\}$, where the electric field establishes itself self-consistently so that there is no local charge accumulation on the time average at the dielectric surface, there are no significant fringing fields and the field corresponds to the TEM mode supplied through a coaxial cable to the reactor.

The dc self-bias sustained at the blocking capacitor is determined from  Kirchhoff's laws for the entire circuit consisting of the rfMS reactor and the external network (see \cite{Eremin2022} for a treatment of a more general external network in the energy-conserving implicit PIC algorithm). It follows that
\begin{equation}
\int\limits_0^{\xi_{p1}} \left(\frac{\epsilon_0 rJ_r }{J_z}\left(\frac{\partial E_\eta}{\partial t}\right)^{n+1/2}+j^{\eta,n+1/2}\right){\rm d}\xi = \frac{C(V^{n+1}-\phi_0^{n+1})-q^n}{\Delta t} \label{eq2_3}
\end{equation}
with $E_\eta$ the electric field's covariant axial component sampled at the powered electrode from the reactor's side, $V$ the generator voltage source, $j^\eta$ the axial contravariant component of the plasma current (see Eq.(\ref{eq2_3})) at the electrode, and $q^n$ the capacitor charge at the previous time level. Although the precise value of $C$ is not important, it has
to be large enough to reduce the potential oscillations due to the charged particles impinging on the powered electrode from the plasma.
In the considered simulations the value of $C$ was chosen to be $3$ nF.

One can then calculate $\phi^{n+1}$ using splitting of the electric field at the time level $n+1$ into a ``vacuum'' part $E^v$ with the unity Dirichlet boundary condition for the corresponding potential:
\begin{equation}
	\cases{
		 1 \, , \, {\rm powered \, electrode} \\
		 \ln(r/r_{p2})/\ln(r_{p1}/r_{p2}) \, ,\, \{ r_{p1}\leq r \leq r_{p2} , z=0 \} \\
0 \, , \, {\rm elsewhere} 		 
	}
	\label{eq2_4}
\end{equation}
and zero charge density (note that the ``vacuum electric field'' is then equal to $\phi_0^{n+1}E_\eta^v$), and the ``plasma-induced'' part $E^\rho$ using the zero Dirichlet boundary condition everywhere and the plasma charge density as a source in the Poisson equation \cite{vahedi_1997}. This leads to 
\begin{equation}
\phi_0^{n+1} = \frac{V^{n+1}-\frac{q^n}{C}-\frac{1}{C}\int\limits^{\xi_{p1}}_{0}
\left(\frac{\epsilon_0 rJ_r }{J_z}(E_\eta^\rho-E_\eta^n)+\Delta t j^{\eta,n+1/2}\right){\rm d}\xi}{1 + \frac{\epsilon_0}{C}\int\limits^{\xi_{p1}}_{0}\frac{rJ_r }{J_z}E_\eta^v {\rm d}\xi}  \label{eq2_5}  
\end{equation}
with $E_\eta^n$ the axial electric field at the previous time level $n$.  

The neutral gas is chosen to be argon and the corresponding chemistry set included elastic scattering, ionization, and excitation \cite{phelps_1999} for the electron-neutral collisions and elastic scattering along with the charge exchange \cite{phelps_1994} for the ion-neutral collisions. The collisions were implemented using the null-collision Monte Carlo algorithm \cite{vahedi_1995} modified for GPUs \cite{mertmann_2011} using the Marsaglia xorshift128 pseudorandom number generator \cite{Marsaglia_2003} randomly initialized for each thread. 

In contrast to the dcMS discharges, where ion-induced secondary electrons accelerated by the cathode sheath voltage
are vital for their very sustainment, secondary electrons in rfMS discharges are often not essential and can be neglected (e.g., \cite{zheng_2021,eremin_2021}). Because of this fact and the complexity of the overall electron heating dynamics, we omit such processes from the present numerical model to be able to concentrate on new electron heating/energization mechanisms inherent to the rfMS discharges and to show that they can be well sustained without secondary electron emission.

\begin{figure}[h]
\centering
\includegraphics[width=8cm]{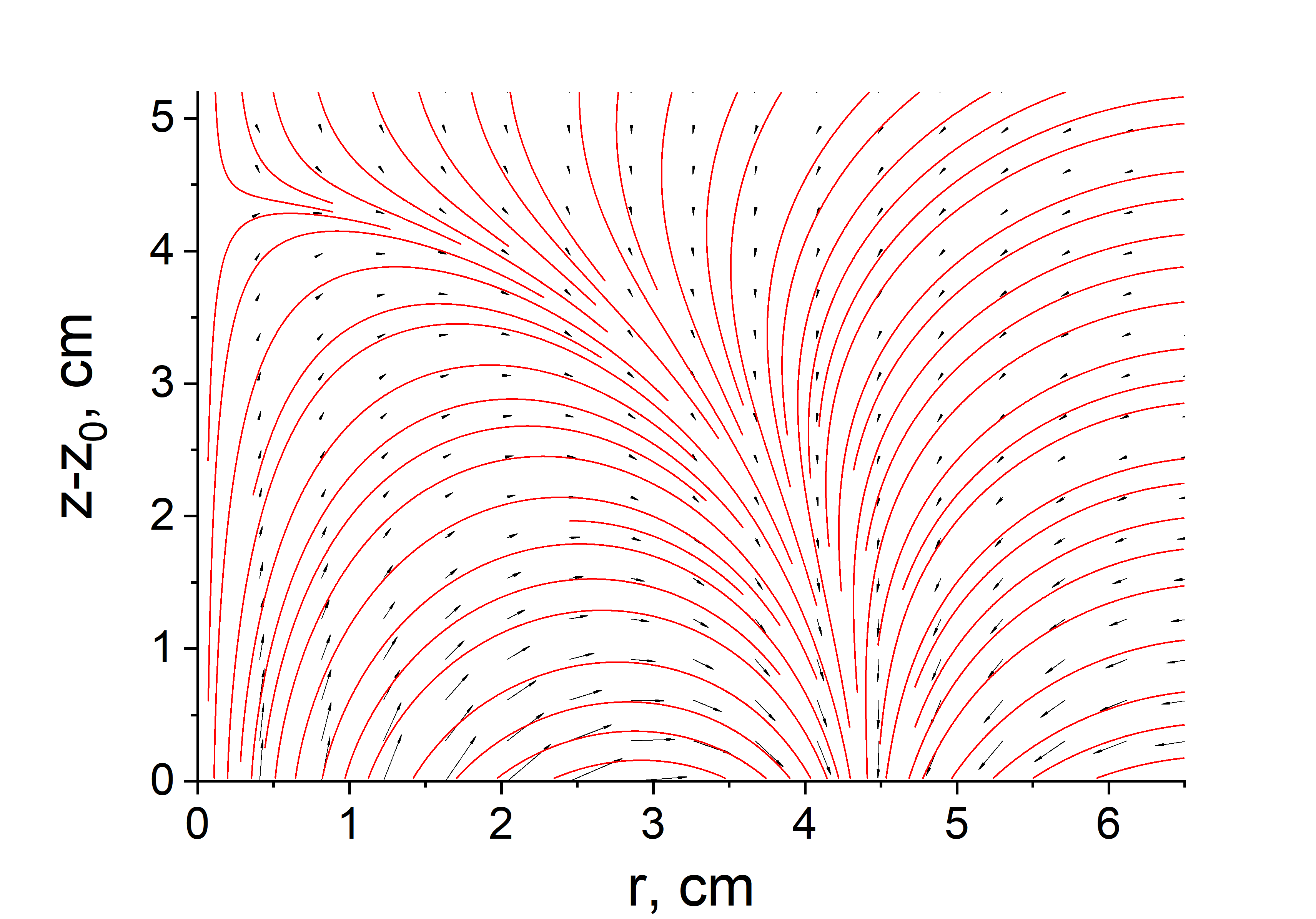}
\caption{Topology of the planar magnetron magnetic field adopted in the present work: magnetic field lines (red lines) and vectors indicating the local
magnetic field direction (black arrows).}
\label{magn_field_geom}
\end{figure}

The magnetic field is chosen to model the field relevant to the experimental device described in \cite{oberberg_2019,oberberg_2018,oberberg_2020} and is approximated using the vacuum solution series similar to \cite{kadlec_1995}. 
The corresponding coefficients are given in \ref{Appendix_a} and it is assumed that due to the low kinetic pressure of plasma in the region where it is well magnetized, the magnetic field modifications due to the plasma currents were negligible. The corresponding magnetic field topology is shown in 
Fig.\ref{magn_field_geom}, both in terms of magnetic lines and vectors indicating the local magnetic field direction. To keep further plots cleaner, in the following we will use only the latter to give information on the magnetic field.


In the following section we analyze results of a numerical simulation with a conducting target and a sinusoidal voltage with a frequency of $13.56$ MHz and an amplitude of $900$ V in argon at $0.5$ Pa. Due to discharge asymmetry the self-bias remained at a large negative value of $-827$ V. The numerical parameters were chosen as follows: In the logical coordinates we used a uniform grid with $(161 \times 258)$ cells in the radial and the axial directions, the time step $2.5 \times 10^{-11}$ s, and the number of superparticles per cell in the converged state was around 500 for the plasma-filled cells.

\section{Results} \label{sec3}


\subsection{Electron transport parallel and perpendicular to the magnetic field \label{ss3a}}

\begin{figure}[h]
\centering
\includegraphics[width=14cm]{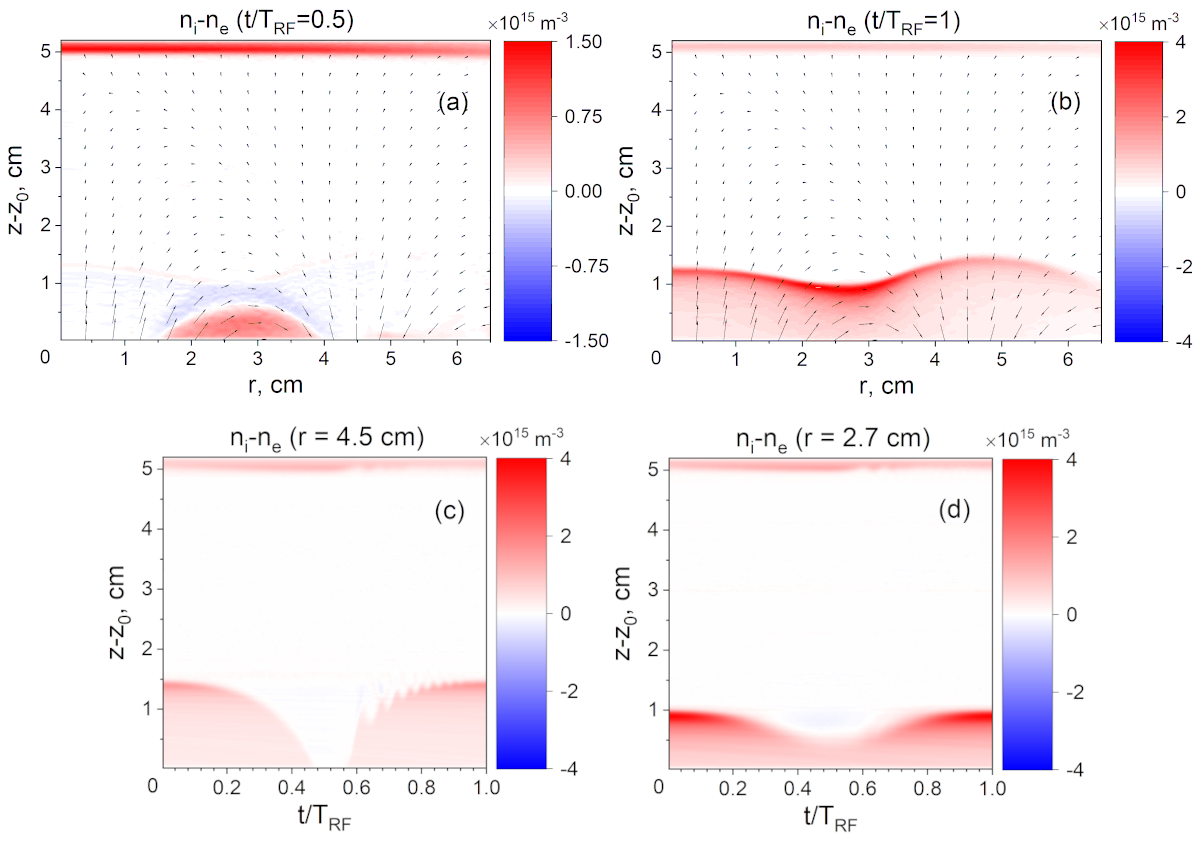}
\caption{Charge density evolution in the main chamber: two-dimensional snapshots in the $(r,z)$ plane taken at the moments of the collapsed sheath (a) and the expanded sheath (b), and the charge density (axial) spatio-temporal evolution at a radial location away from the racetrack region (c) and at the racetrack location (d).}
\label{Fig2}
\end{figure}

As argued above, the electron power absorption dynamics is closely related to the electron transport. Therefore, it is important to understand how electrons move under the influence of the electric and the magnetic fields. Drawing from the knowledge base of the electron dynamics in conventional unmagnetized CCRF discharges, one can assume that the plasma bulk has to be quasineutral, while close to the electrodes there emerge time-modulated electron-depleted areas accommodating most of the discharge voltage drop. Such areas are known as plasma sheaths. During the expansion phase they accelerate electrons to large energies in the direction of the plasma bulk, while during the collapse phase electrons flow to the powered electrode. In the case a blocking capacitor is present in the external circuit and no DC current is possible, the electron flux to the powered electrode has to be sufficient to neutralize the average positive charge brought to the electrode by ions. The magnetic field introduces a strong anisotropy in the behavior of magnetized electrons. Their flow across the magnetic field becomes impeded, whereas their motion along the magnetic field lines is very similar to the unmagnetized case (in this case the magnetic field influences the electron motion parallel to the magnetic field only through the mirror force \cite{keidar_2005} arising in a nonuniform magnetic field).
Combined with the typical magnetic field topology of a typical planar magnetron, the anisotropy in electron flow leads to a radial nonuniformity in the sheath behavior. This can indeed be observed in Fig.\ref{Fig2}, where the difference in ion and electron densities is depicted during the sheath collapse (Fig.\ref{Fig2}a) and when the sheath is expanded (Fig.\ref{Fig2}b). In Fig.\ref{Fig2}a one can clearly see that the magnetic field strongly inhibits the electron motion around the racetrack area, where electrons would experience a substantial magnetic field component orthogonal to their motion on their direct path to the powered electrode. Rather, they prefer to take the ``path of least resistance'' and move to the powered electrode along the magnetic field lines once they arrive at such a magnetic field line connecting them to the powered electrode. Therefore, the sheath collapse occurs only in the areas close to the electrode center and the electrode edge, whereas the sheath width above the racetrack is modulated much more weakly. As one can see in Fig.\ref{Fig2}a, the resulting charge density profile has the form of a positively charged bump above the racetrack. Since the bump's form for the sheath above the racetrack at this moment is a result of the electrons' magnetization, we will call it ``magnetized sheath''. The observed electron density enhancement immediately above the bump (the negative ``halo'') indicates that there should be some resistance to electron motion even along the magnetic field leading to the electron accumulation there. It will be later argued that such a resistance is due to the mirror force (see \ref{ss3b}). The largest sheath width oscillation's amplitude occurs away from the racetrack region (see Fig.\ref{Fig2}b), where the entire sheath width's time modulation very much resembles its counterpart in an unmagnetized CCRF discharge, as Fig.\ref{Fig2}c shows. One can observe that the positively charged sheath fully collapses during $0.46 < t/T_{RF} < 0.57$, and thereafter it rapidly expands, the edge exhibiting plasma series resonance (PSR)-like oscillations \cite{annaratone_1995,klick_1996,czarnetzki_2006,wilczek_2016} along the magnetic field lines, which leads to the NERH. In contrast, Fig.\ref{Fig2}d demonstrates that the sheath above the racetrack oscillates with a much smaller amplitude and never collapses fully. Note that it also does not show PSR oscillations of the expanding edge.    

\begin{figure}[h]
\centering
\includegraphics[width=12cm]{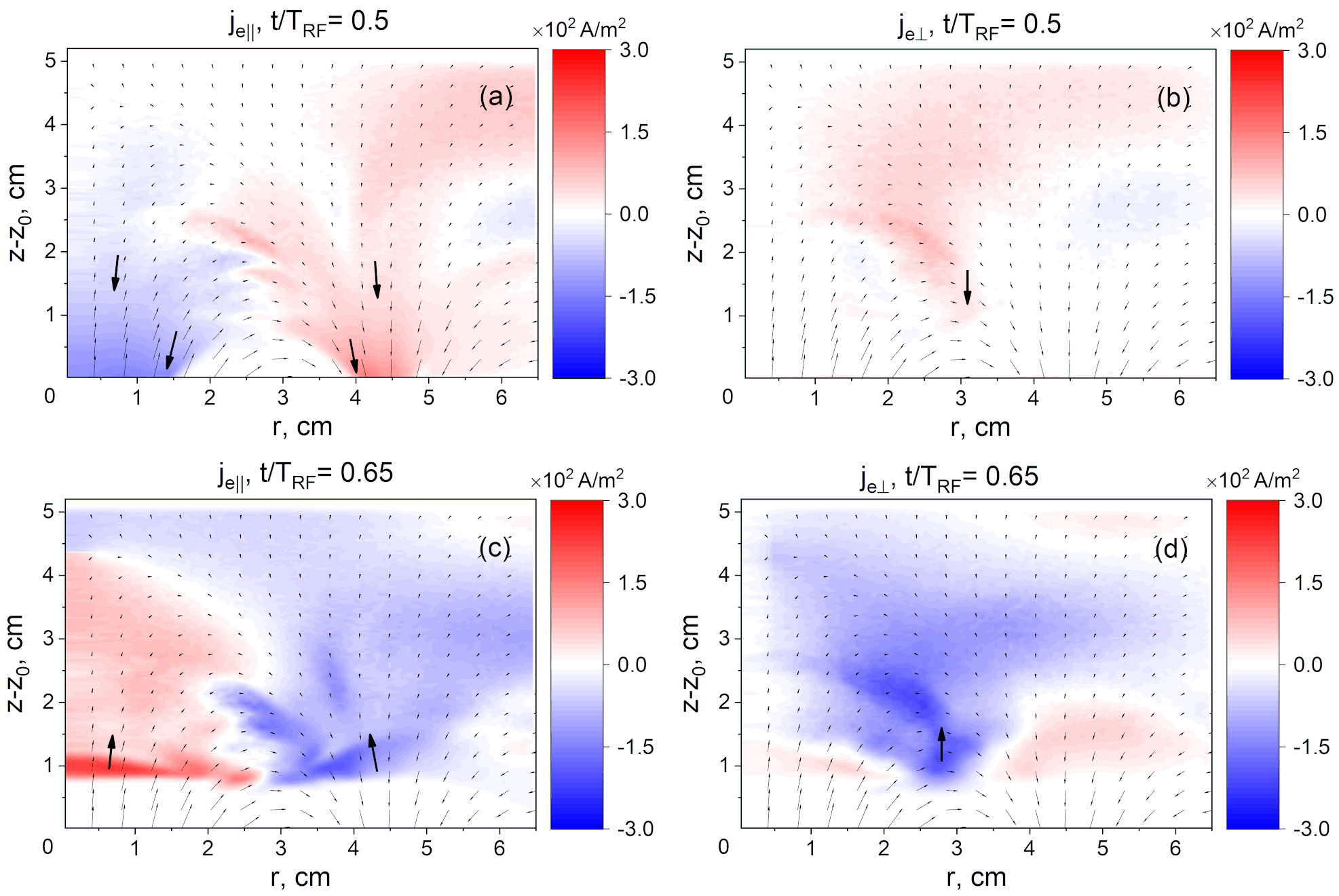}
\caption{Longitudinal and transverse electron current densities sampled at the collapsed sheath [(a) and (b)], and at the sheath expansion [(c) and (d)] phases. The decomposition into the transverse and longitudinal components has
been performed following Eq.(\ref{Eq3_1}). Note that the magnetic field close to the target has an arc-like shape and tends to point from the target at the target electrode's radial center and towards the target at the periphery of the
target electrode. Hence the formal change of sign (a) and (c), although in the lab coordinates the corresponding component has similar directions indicated by the black arrows.}
\label{Fig3}
\end{figure}

It is well known that classical collisional transport coefficients of magnetized charged particles can be very different parallel and perpendicular to the magnetic field lines. For example, the ratio of the classical (collision-dominated) diffusion coefficients parallel and perpendicular to the magnetic field is approximately proportional to $\omega_{ce}^2/\nu_m^2$ with $\omega_{ce}$ the Larmor frequency and $\nu_m$ the momentum relaxation frequency \cite{lieberman_2005}. An extreme case would be a class of highly magnetized plasmas such as those produced in fusion experiments, where the longitudinal and transverse transport coefficients can differ by many orders of magnitude. Although in the case considered here the difference is not that large, it is substantial for a large part of the discharge where electrons can be considered magnetized, warranting a decomposition of vector quantities related to the discharge dynamics into the longitudinal and transverse parts. This can be easily performed by using the following expressions,
\begin{equation}
\begin{array}{lll}    
{\bf v} = v_\parallel {\bf b} + {\bf \tilde{v}_\perp} \\
v_\parallel = ({\bf v}\cdot{\bf b}) \\
{\bf \tilde{v}_\perp} = {\bf v_\perp} + v_\theta{\bf e_\theta} \\
v_\theta = ({\bf v}\cdot{\bf e_\theta}) 
\end{array} \label{Eq3_1}    
\end{equation}
where ${\bf v}$ is an arbitrary vector and ${\bf b} = {\bf B}/B$ is a unit vector pointing in the magnetic field direction. Here we have explicitly used the fact that ${\bf b}$ is located in the $(r,z)$ plane, so that ${\bf v_\perp}$ denotes the transverse part in $(r,z)$ plane, and we use the tilde when the azimuthal component is also included. Note that in this decomposition the sign of $v_\parallel$ is positive when ${\bf v}$ is co-directed with the local magnetic field and is negative when these vectors are counter-directed.

Application of Eqs.(\ref{Eq3_1}) to the current density vector at the moment of the full sheath collapse, $t/T_{RF}=0.5$, and around the moment of its fastest expansion phase, $t/T_{RF}=0.65$, confirms that during the sheath collapse one of the main electron current channels goes perpendicular to the magnetic field approximately above the racetrack (see Fig.\ref{Fig3}b) and then parallel to the magnetic field lines for $z<1$ cm (indicated by the lower arrows in Fig.\ref{Fig3}a). This can be explained as follows: the majority of the electron population is well confined in the EMCR by the magnetic field in the direction perpendicular to the magnetic field, and by the sheath's negative electrostatic potential or the magnetic mirror effect in the direction parallel to the magnetic field. During the sheath retraction those of the electrons from the EMCR, which were not confined by the mirror effect, will follow the retracting sheath along the boundary of the magnetized sheath. However, their motion can still be impeded by the mirror force on that path later, because the magnetic field strength increases as they come closer to the target. Therefore, a reversed electric field builds up along the bump-like boundary of the magnetized sheath, which increases the longitudinal kinetic energy of such electrons, and thus decreases their pitch angle (formed between the velocity and the local magnetic field vectors), so that they could continue their flow to the powered electrode. Furthermore, Fig.\ref{Fig3}a exhibits two additional longitudinal electron transport channels (indicated by the upper arrows) along the magnetic field lines on both sides of the EMCR and the magnetized sheath. Electrons move in these channels predominantly along the magnetic field lines, thus they are affected by the magnetic field only through the mirror force.
During the sheath expansion the main electron current channels are again formed perpendicular to the magnetic field above the racetrack region (see Fig.\ref{Fig3}d) and parallel to the magnetic field on its sides (indicated by the arrows in Fig.\ref{Fig3}c). The electron mobility is substantially lowered by the Lorentz force in the EMCR region of the transverse channel compared to the parallel electron mobility reduced by a weaker mirror force in the longitudinal channels.
However, the density of electrons trapped in the EMCR region is significantly higher compared to the other discharge regions, therefore one can expect the electron conductivity in the transverse and the longitudinal channels to be comparable. In this way, one can see that such a planar magnetron rfMS discharge exhibits features of magnetized and unmagnetized CCRF discharges simultaneously.

It is also interesting to observe comb-like structures on the plots of $j_{e\parallel}$ during the sheath collapse and expansion (see Fig.\ref{Fig3}a and Fig.\ref{Fig3}c). It indicates that there is a shear in the electron net longitudinal velocity related to the electrons confined in the EMCR. While such a structure could suggest a wave-like phenomenon, it does not seem to affect the electron power absorption in the corresponding area (discussed later) in any way. Therefore, we do not address it in the present work.   

In contrast to electrons, ions are unmagnetized and move solely under the influence of the electrostatic potential. Their transport in the sheath region is independent from electrons and is therefore non-ambipolar. This is possible due to the shorting currents in the conducting target \cite{simon_1955,lafleur_2012}. 
Since most of the ions are produced in the narrow EMCR above the racetrack, their flux at the powered electrode peaks at the racetrack radial location. However, the total electron and ion charges carried to the powered electrode by the corresponding currents over the rf period are equal.  

\begin{figure}[h]
\centering
\includegraphics[width=14cm]{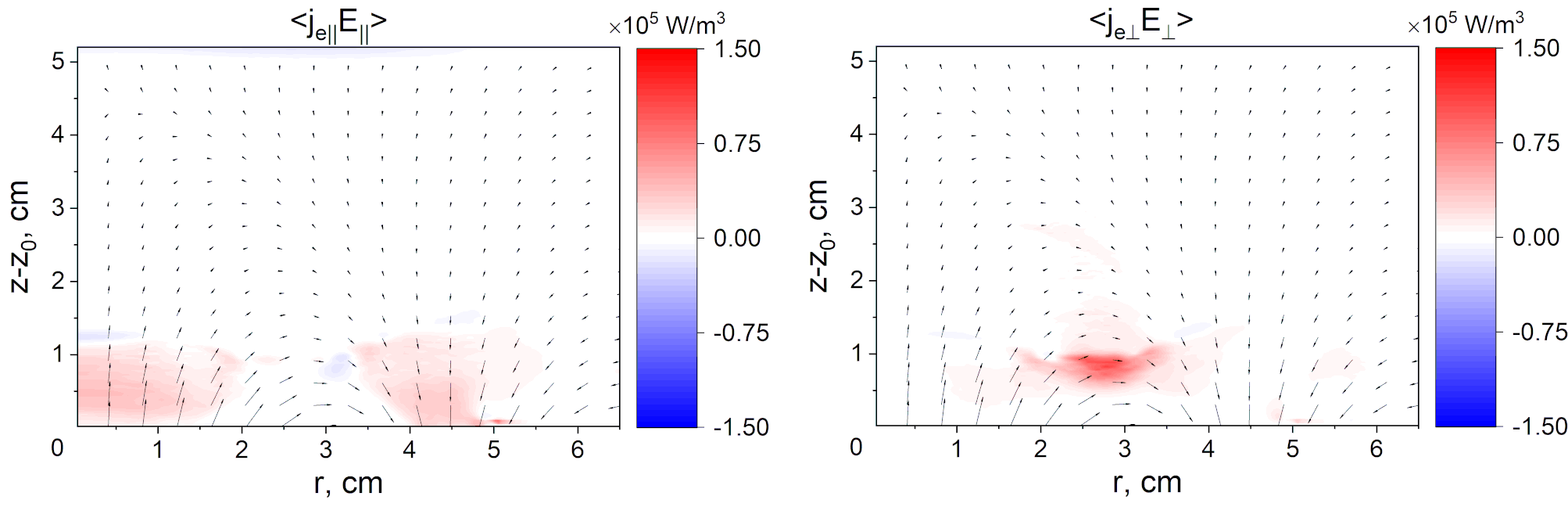}
\caption{Time-averaged longitudinal (left) and transverse (right) power densities absorbed by electrons.}
\label{Fig4}
\end{figure}

Similar decomposition of the time-averaged power absorption shown in Fig.\ref{Fig4} demonstrates that it is dominated by the transverse part, which is concentrated in the EMCR. It coincides with the ``electron transverse transport'' channel mentioned above.
The weaker power absorption along the magnetic field lines takes place away from that region, being most pronounced in the powered sheath region.  
In the following, we consider time-dependent dynamics of the longitudinal and the transverse power absorption during the rf cycle and identify the related electron heating/energization mechanisms.

\subsection{Longitudinal power absorption \label{ss3b}}

For the following discussion it is convenient to divide the radial domain into the following three zones:
in zones 1 (0-1\,cm) and 3 (4-5\,cm) the magnetic field close to the powered electrode is predominantly perpendicular, whereas in zone 2 (2-3.5\, cm) the magnetic field is predominantly parallel to the electrode surface.
The discussion in this section is related to the excitation pattern's axial-temporal dynamics in the zones 1 and 3 (see also the companion work \cite{berger_2021}).

\begin{figure}[h]
\centering
\includegraphics[width=16cm]{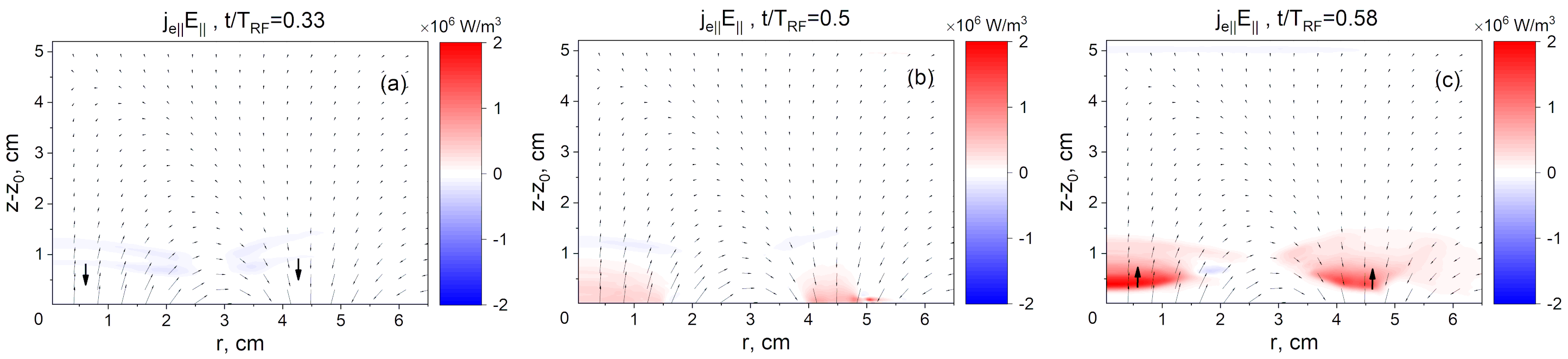}
\caption{Longitudinal power density absorbed by electrons taken at sample moments during the sheath retraction (a), sheath collapse (b), and sheath expansion (c).}
\label{Fig6}
\end{figure}

Fig.\ref{Fig6} shows basic power absorption patterns which can be observed in the longitudinal power absorption profile during the sheath retraction (Fig.\ref{Fig6}a), collapse (Fig.\ref{Fig6}b), and sheath expansion (Fig.\ref{Fig6}c), respectively. On the one hand, there are typical patterns, which can also be seen in unmagnetized CCRF discharges. One can observe the electron cooling (heating) at the stationary sheath-presheath boundary related to the ambipolar electric field \cite{schulze_2015}
and at the edge of the retracting (expanding) sheath (shown by the arrows). On the other hand, Fig.\ref{Fig6}b exhibits a new electron heating mechanism, which is due to an enhanced reversed electric field needed to overcome the mirror force $-\mu\nabla_\parallel B$, with $\mu = \varepsilon_\perp/2B$ the magnetic moment and $\varepsilon_\perp$ the Larmor rotation energy defined in Eq.(\ref{eq3_2}) under neglect of the electron drift caused by magnetic field nonuniformities. 

\begin{figure}[h]
\centering
\includegraphics[width=14cm]{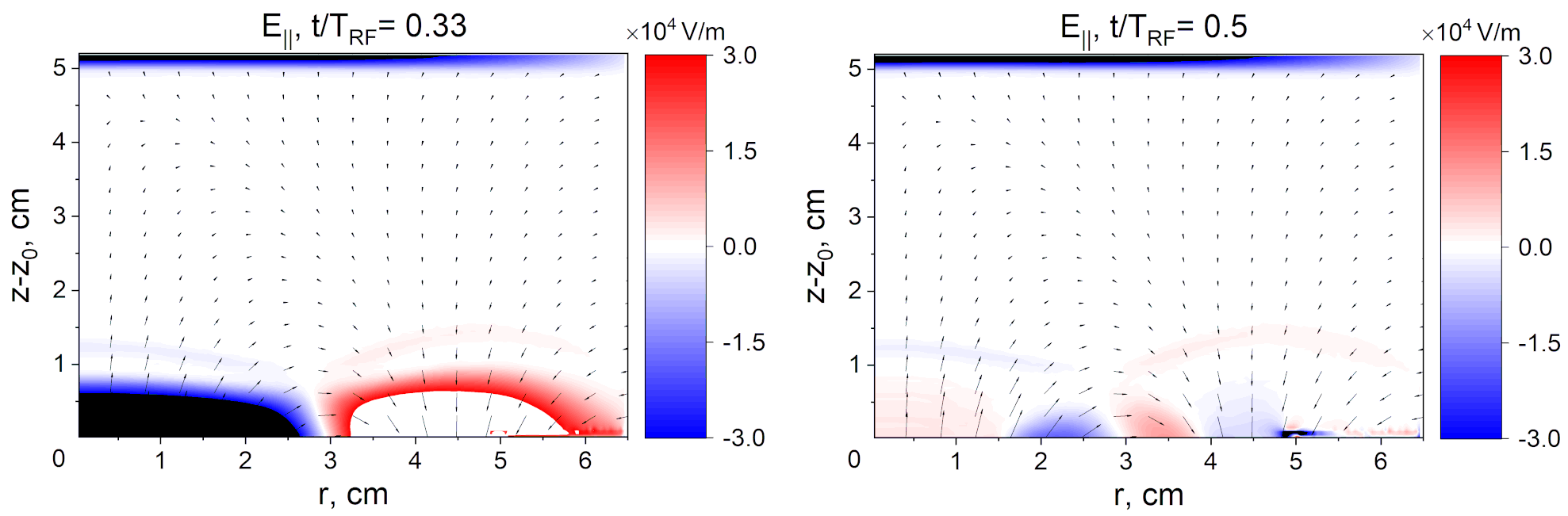}
\caption{Longitudinal electric field taken at sample moments during the beginning of the sheath retraction (left) and during the sheath collapse (right).}
\label{Fig7}
\end{figure}

Fig.\ref{Fig7} compares the longitudinal electric field at the beginning of the powered sheath retraction (left plot) and during the powered sheath full collapse (right plot). Note the electric field reversal during the sheath collapse in the radial target segments ($0<r<1.5\,{\rm cm}$) and ($4\,{\rm cm}<r<5\,{\rm cm}$), where electrons flow to the target along the magnetic field lines (see the discussion in \ref{ss3a}). At the same time, the electric field does not change its ``normal'' negative polarity in the region close to the racetrack, which is due to the fact that the sheath does not collapse there and remains positively charged. The reversed electric field leads to an enhanced power deposition close to the target and is needed to ensure a sufficiently large flow of electrons to the powered 
electrode, which is required to balance the charge deposited by ions during the rf period. At a pressure of $p=0.5$ Pa, the considered collisions cannot be a reason for the electric field reversal in contrast to highly collisional cases \cite{tochikubo_1992,schulze_2008,eremin_2015} and the inertial electric field reversal \cite{schulze_2008,sato_1990,vender_1992} is normally not observed in argon under these conditions. Since electrons predominantly flow along the magnetic lines in this region, the Lorentz force cannot cause the field reversal either. Hence, we conclude that the observed electric field reversal is rather caused by the need to overcome the mirror force related to the magnetic moment's conservation and acting on electrons along the magnetic lines.

\begin{figure}[h]
\centering
\includegraphics[width=12cm]{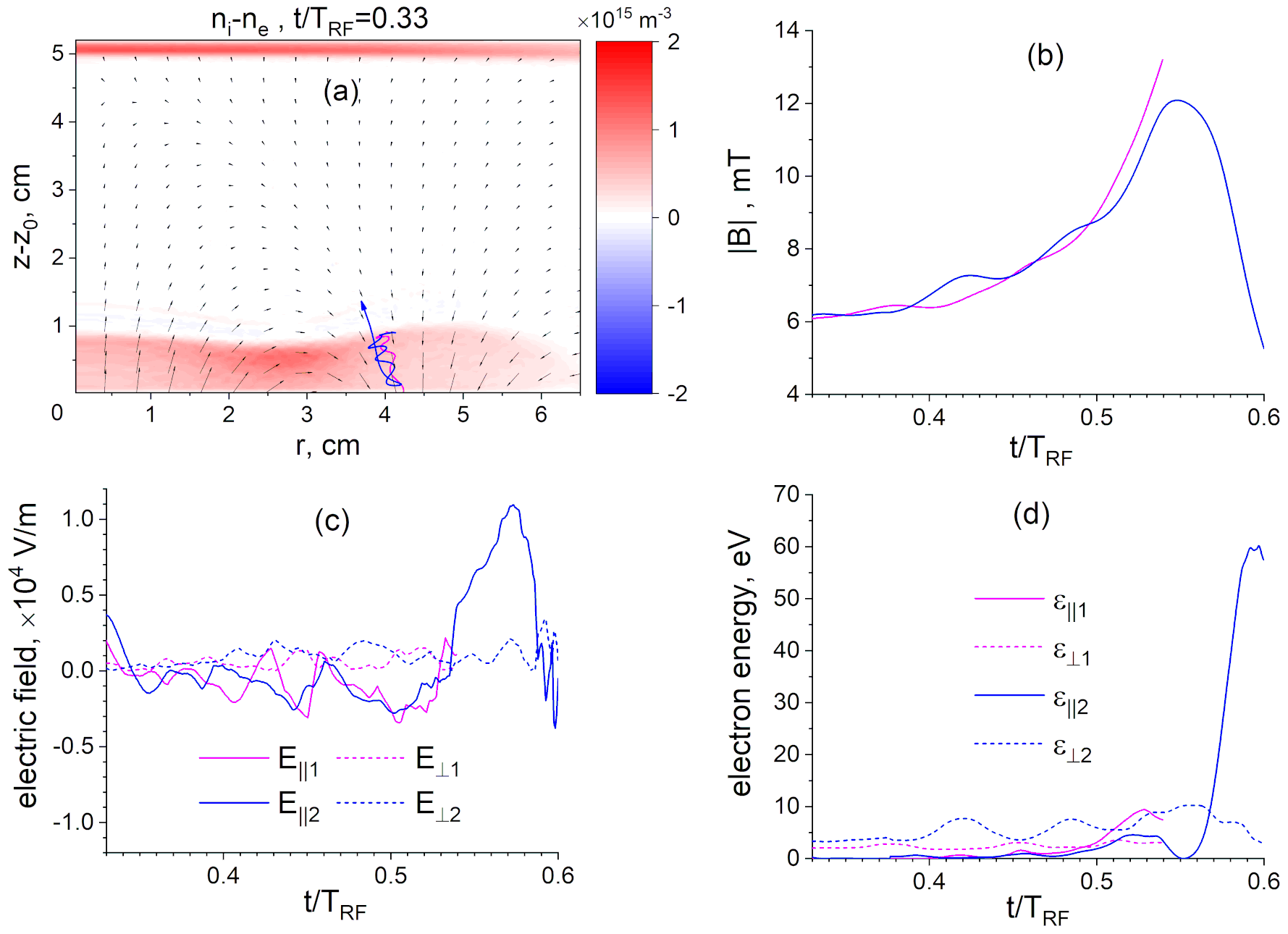}
\caption{Properties of two sample test particles following the powered sheath retraction: their trajectories (a), the experienced magnetic field flux density (b), the experienced longitudinal and transverse electric field components (c), and the kinetic energy of the longitudinal and transverse motion (d).}
\label{Fig7a}
\end{figure}

To support this point, Fig.\ref{Fig7a} follows orbits (Fig.\ref{Fig7a}a) along with the magnetic flux density (Fig.\ref{Fig7a}b), electric field (Fig.\ref{Fig7a}c) and kinetic energies of the motion parallel and perpendicular to the magnetic field lines (Fig.\ref{Fig7a}d) experienced by two test particles during their motion. The particles were selected from a larger number of test particles seeded as a zero-weight copy of random actual simulation particles located close to the edge of the retracting powered sheath when it is still relatively wide ($t/T_{RF}=0.33$, the charge density at this moment is shown for reference). Then, two representative orbits were chosen, which showed typical cases when 
a particle either went down all the way to the powered electrode and got absorbed there (particle 1, magenta color) or until a particle could not reach the powered electrode due to the mirror force acting in the direction away from the powered electrode and 
became accelerated by the expanding powered sheath in the plasma bulk's direction afterwards (particle 2, blue color). Fig.\ref{Fig7a}b demonstrates that the particles experience significant growth of the magnetic flux density as they approach the target located at the powered electrode, so that there should be a substantial mirror force acting in the direction opposite to their motion in this phase. The kinetic energies are calculated as $\varepsilon_{\parallel} = m_e v_\parallel^2 / 2$ and
\begin{equation}
\varepsilon_\perp = \frac{m_e}{2} \left(v_\perp^2+\left(v_\theta-\frac{E_\perp}{B}\right)^2\right)
\label{eq3_2}
\end{equation}
with $v_\perp=|{\bf v_\perp}|$ calculated using Eq.(\ref{Eq3_1}), and $B$ and $E$ taken from Fig.\ref{Fig7a}b and Fig.\ref{Fig7a}c, respectively. Since for the magnetic moment only the rotational velocity matters, 
the ${\bf E\times B}$ drift is subtracted from the azimuthal velocity and it is assumed that the azimuthal drifts due to magnetic field nonuniformities are small. Fig.\ref{Fig7a}d shows that at the start the kinetic energy of the transverse motion is larger than that of the longitudinal one for both particles, indicating their relatively large pitch angles. However, the reversed longitudinal electric field (Fig.\ref{Fig7a}c) substantially increases the longitudinal kinetic energy, whereas the transverse kinetic energy remains limited, so that the pitch angle of both particles effectively decreases (this is also indicated by the decrease in their orbits' number of turns per unit length observed in Fig.\ref{Fig7a}a). For the first particle the pitch angle decrease is sufficient to let the particle reach the powered electrode. In contrast, the second particle could not decrease its pitch angle enough to 
get to the electrode and so remains trapped. In this case the powered sheath expansion and the concomitant change of the electric field polarity, which starts at $t/T_{RF}=0.54$ (Fig.\ref{Fig7a}c), causes this particle to first decrease the longitudinal energy and then to increase it to a large value around $60$ eV due to the NERH energization to be addressed next. The considered electrons attain maximum energy associated with the electric field reversal heating approximately at the moment when the longitudinal electric field changes its polarity, which signifies the transition of the powered sheath motion phase from the collapse to the expansion.

\begin{figure}[h]
\centering
\includegraphics[width=16cm]{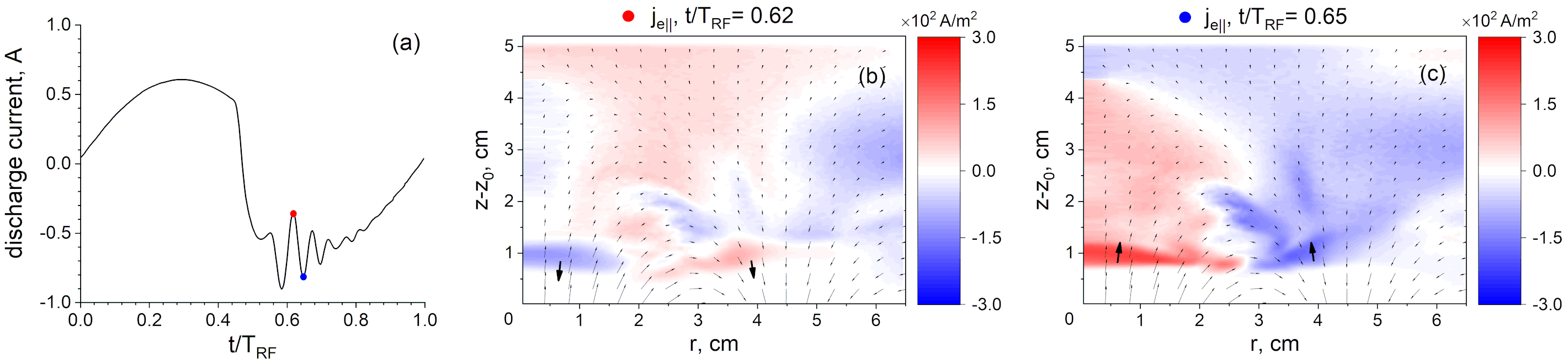}
\caption{PSR oscillations in the longitudinal electron current channels: time evolution of the total discharge current (a), and the longitudinal electron current density sampled at the moments of the first local peak of the total discharge current (b) and the second trough (c), indicated by the red and blue dots.}
\label{Fig8}
\end{figure}

The sheath expansion away from the racetrack region is quite violent and 
is accompanied by strong PSR-like oscillations (Fig.\ref{Fig2}c), whereas time modulation of the magnetized sheath width is relatively weak (Fig.\ref{Fig2}d).
The excitation of the PSR is also manifested in oscillations of the total discharge current shown in Fig.\ref{Fig8}a. As argued above, the PSR is triggered in the longitudinal electron current channels on the sides of the racetrack region 
, whereas the absence of the PSR signature on the magnetized sheath oscillations above the racetrack region (Fig.\ref{Fig2}d) indicates that the transverse current channel does not exhibit a significant PSR excitation at least for the conditions considered. This can be explained as follows: for the electron dynamics along the magnetic field lines the magnetic field's influence is weakened, therefore the dynamics resembles that of an unmagnetized CCRF discharge. The mirror force acting along the field lines does not have much influence on the electron motion in the longitudinal channels during the sheath expansion because the energy of their longitudinal motion acquired due to the electron acceleration by the sheath expansion is much larger than the transverse kinetic energy, so that their pitch angle is small.
However, the magnetic field strongly affects the transverse electron dynamics above the racetrack, which significantly changes the resonant behavior of the corresponding current channel \cite{oberberg_2019,wang_2021,liu_2021}. The lack of the PSR excitation in the transverse current channel observed here can be attributed to the shift of the corresponding resonance frequency caused by the magnetic field (see the companion paper \cite{eremin_2021}).
One can additionally observe the connection between the PSR oscillations and the longitudinal current dynamics in Fig.\ref{Fig8}b and Fig.\ref{Fig8}c, which show the longitudinal current density during the first major peak and the second major trough in the total current (Fig.\ref{Fig8}a) after the PSR excitation. Directions of the mean electron drift at the respective moments are indicated by
the arrows. One can observe the longitudinal electron current density reversal due to the PSR oscillations. 

\begin{figure}[h]
\centering
\includegraphics[width=16cm]{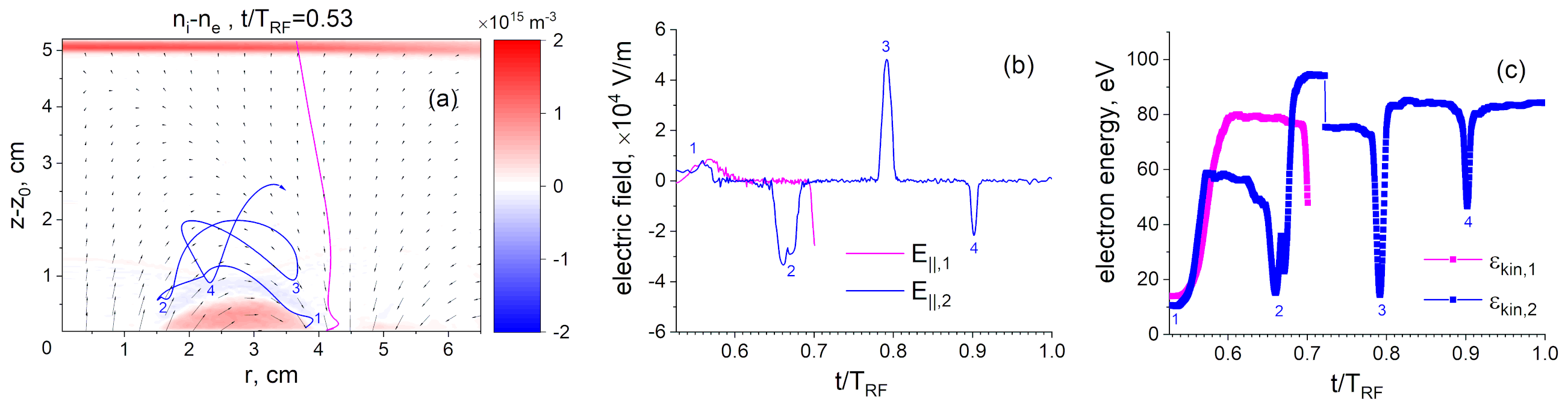}
\caption{Properties of two sample test particles energized due to the NERH and MBH: trajectories (a), the longitudinal electric field components experienced by the particles (b), and their kinetic energies (c).}
\label{Fig8a}
\end{figure}

The PSR oscillations cause a rapid motion of the powered sheath edge, which leads to a drastic power absorption enhancement \cite{mussenbrock_2008}. During the sheath edge expansion along the magnetic field lines electrons are repelled from the sheath by its negative potential, which results in the generation of highly energetic electrons with kinetic energies of $50$ eV or more for the chosen parameters. 
The magnetization parameter estimated as the ratio of the Larmor radius to the magnetic field scale is not necessarily small for them, so that the simple representation of their motion as a cycloid winding around a guiding magnetic field line may no longer be valid.
Rather, their accessibility domain should be obtained based on the effective potential resulting from the canonical azimuthal momentum's conservation in an axisymmetric device \cite{wendt_1988,sheridan_1989}. Such an analysis shows that energetic electrons can easily detach from the magnetic field lines \cite{sheridan_1989} and become virtually unmagnetized. Fig.\ref{Fig8a}a shows  orbits of two different test particles started to be traced at a moment of time before the powered sheath expansion begins at $t/T_{RF}=0.53$ (for the reference, the charge density indicating the sheath width at that moment is also plotted). They rapidly reach comparable energies
due to the acceleration by the longitudinal electric field (see Fig.\ref{Fig8a}b)
during the sheath expansion (see Fig.\ref{Fig8a}c). However, depending on their initial positions and velocity components, they take different paths. 

The first particle (magenta color) gains large longitudinal energy during the sheath expansion. The effective potential resulting from the magnetic and electric field is too small to confine the particle within the discharge, so that after traversing the discharge practically along the field lines
the particle is lost at $t/T_{RF}=0.7$. The particle energy is so high that the grounded sheath electrostatic potential was not capable of reflecting it back into the discharge (Fig.\ref{Fig8a}c). Note that such a rapid loss of the particle to the reactor walls or electrodes before utilizing its energy for particle-producing ionization collisions or excitation events detected in the experiment means that the contribution of the longitudinal heating/energization dynamics can be expected to have a limited impact on the time-averaged excitation or ionization profile.

The second particle with the trajectory marked by the blue color also starts off by acquiring a large energy (about $60$ eV) by interaction with the rapidly expanding sheath (1). However, due to different initial conditions from the first particle, its trajectory is bent by the magnetic field, so that it follows an arcing magnetic field line, which guides it to collide with the expanding sheath on the other side of the racetrack region (2), which increases its energy even more (up to $87$ eV). It is interesting to see that during the interaction with the expanding sheath the particle experienced an elastic ``backscattering'' collision, which caused it to collide with the expanding sheath once more, contributing to a large ``secondary'' energization similar to unmagnetized CCRF discharges \cite{schulze_2015}. Then, the particle is guided by the magnetic field once again to the opposite side of the racetrack region, where it experiences another interaction with the expanding sheath (3), which re-energizes it after the energy loss due to an inelastic collision that happened at $t/T_{RF}=0.72$ (Fig.\ref{Fig8a}c). Note that the observed dynamics suggests another new electron heating/energization mechanism, which is possible only due to the arc-like magnetic field guiding electrons between collisions with the expanding sheath on the opposite sides of the racetrack region. This resembles the bounce-heating mechanism in unmagnetized CCRF (e.g., \cite{kaganovich_1996,liu_2011}), albeit here the particle collides with the sheaths not at different electrodes, but with the same powered sheath. Using this analogy, we propose to call the corresponding heating mechanism ``magnetized bounce heating'' (MBH). Note, however, that the MBH is different from the bounce-resonance heating mechanism investigated in those references in that it is non-resonant and requires only that the bounce period should be substantially smaller than the rf period, so that the electron could collide with the sheath during the same expansion phase. Consequently, this phenomenon resembles the non-resonant bouncing of the energetic electron beam in an unmagnetized plasma, which was observed in \cite{schulze_2008}.
Since the MBH is caused by the longitudinal dynamics, this mechanism is different from the MSH proposed in \cite{lieberman_1991}, which is due to the Larmor gyration, i.e. the transverse dynamics. The MBH could be comparable to or stronger than the MSH as long as the bounce frequency is larger then the cyclotron frequency, which is possible due to the strong initial heating by the NERH. Another reason why the MBH mechanism can be stronger than MSH is that the longitudinal sheath motion exhibits strong PSR oscillations (see Fig.\ref{Fig2}c) leading to large energy gains in the MBH, whereas the magnetized sheath moves smoothly and has a smaller modulation amplitude (see Fig.\ref{Fig2}d). Note that during the considered time interval the particle experiences another collision with the sheath region (4), this time with the magnetized sheath in the racetrack region. In this case it satisfies the premises of the MSH. However, since in this case the sheath is almost fully expanded and its expansion velocity is small, there is no significant energy increase.


\subsection{Transverse power absorption \label{ss3c}}

\begin{figure}[h]
\centering
\includegraphics[width=16cm]{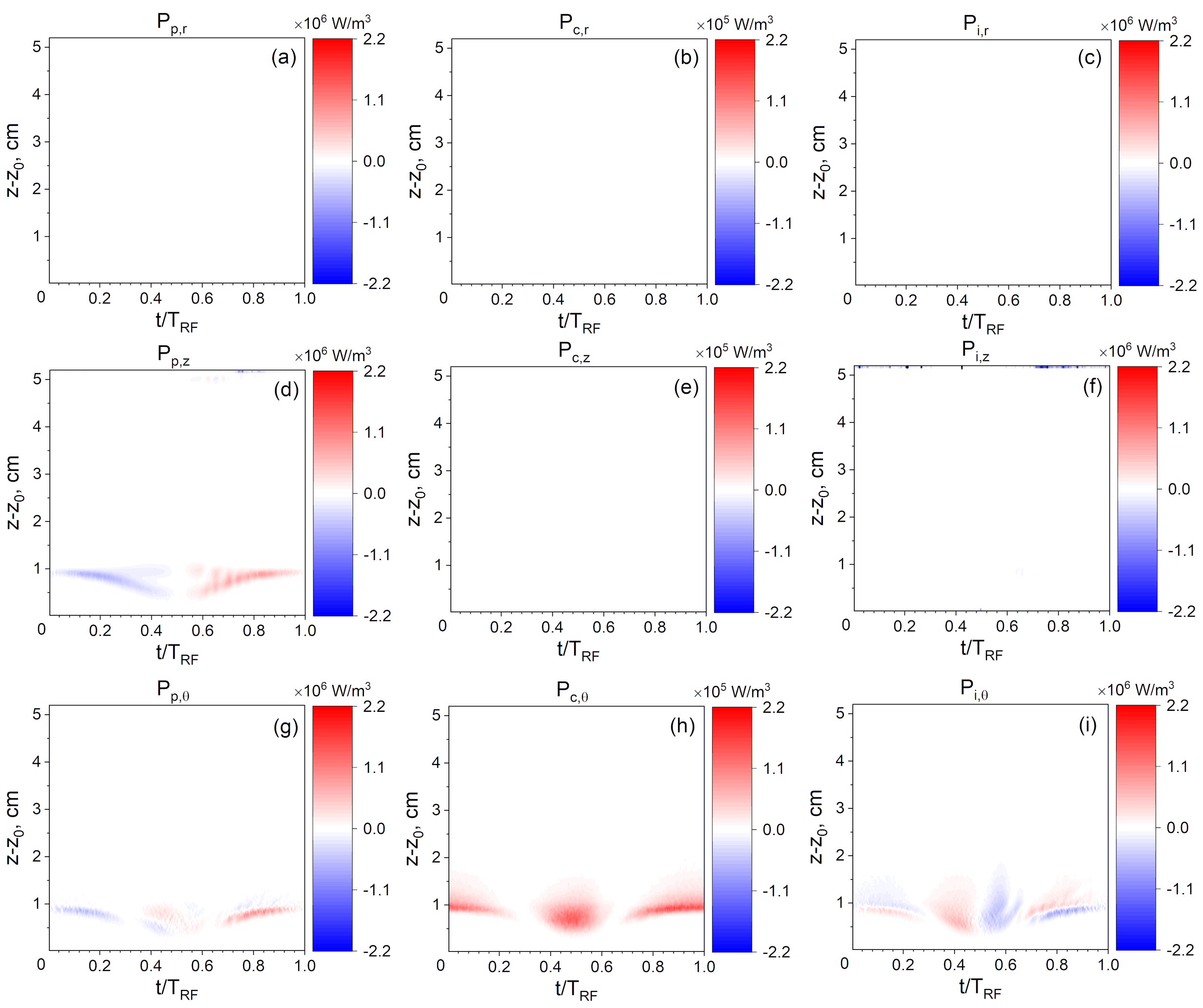}
\caption{Temporal evolution of the axial profiles of the mechanical energy balance terms sampled at the radial racetrack location $r=2.7$ cm.}
\label{Fig10}
\end{figure}

The following discussion is related to the excitation pattern's axial-temporal dynamics in the zone 2 (see the companion work \cite{berger_2021}).
As can be seen in Fig.\ref{Fig4}, right, the transverse power absorption in the EMCR yields
the dominant net contribution over the rf period. 
It is interesting to analyze its dynamics within the rf period. 
Fig.\ref{Fig10} shows different terms in the mechanical energy balance decomposition of the power absorption (see 
\ref{Appendix_b}) sampled in time and along the axial coordinate at a fixed radial location in the racetrack region, $r=2.7$ cm. Despite the azimuthal electric field being assumed to be zero due to the azimuthal symmetry,
the azimuthal terms in the mechanical balance equation arise from the fact that   
the power gained from the electric fields in the $(r,z)$ plane and described by $j_rE_r + j_zE_z$ is channeled by the Hall terms to the azimuthal direction \cite{zheng_2021,eremin_2021,zheng_2019} (see also \ref{Appendix_b}), reflecting the energy flow related to the Hall heating mechanism (see the Introduction). According to \cite{eremin_2021}, the transverse electric field combined with the magnetic field generates a strong azimuthal drift with the total average kinetic energy in the inelastic range. It can be seen in Fig.\ref{Fig10} that this process is essentially collisionless, since the ``collisionless'' energy transfer rates $P_{p\theta}$ and $P_{i\theta}$ are substantially greater than the collisional $P_{c\theta}$. The large terms $P_{p\theta}$ and $P_{i\theta}$ reflect the fact that the time-dependent electric field ``pumps'' energy in the azimuthal ${\bf E \times B}$ drift motion during the transverse electric field's increase, and ``absorbs'' this energy back during the transverse electric field's decrease \cite{eremin_2021}. 


\begin{figure}[h]
\centering
\includegraphics[width=8cm]{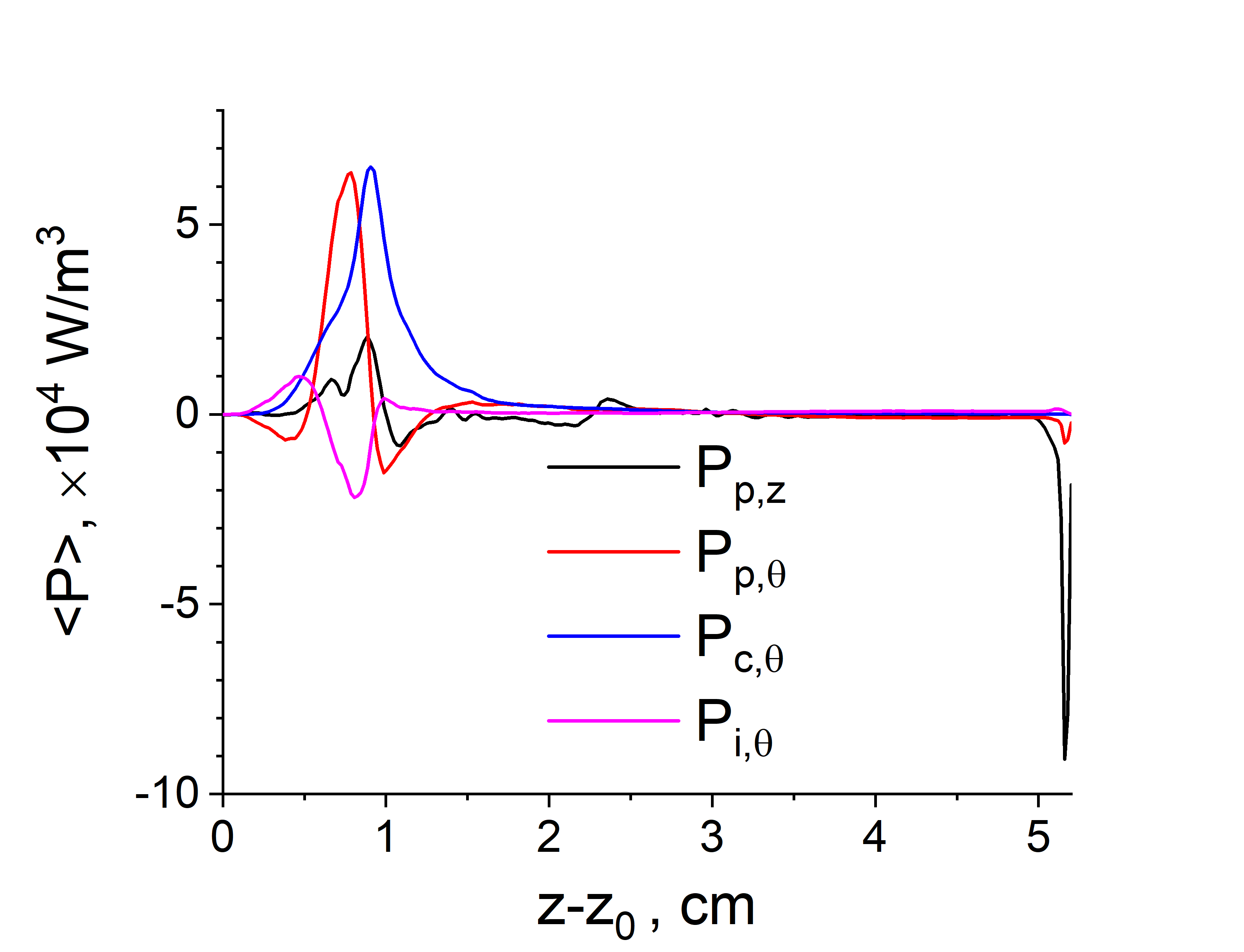}
\caption{Period-averaged axial profiles of $P_{p,z}$, $P_{p,\theta}$, $P_{c,\theta}$, and $P_{i,\theta}$ sampled at the radial racetrack location $r=2.7$ cm.}
\label{Fig10c}
\end{figure}

The only significant non-azimuthal contribution to the transverse power absorption at the racetrack's radial location comes from the ``pressure heating'' component $P_{p,z}$ shown in Fig.\ref{Fig10}d, which shows positive (negative) energy exchange between electrons and the electric field during the sheath expansion (retreat). The oscillations of $P_{p,z}$ in Fig.\ref{Fig10}d during the sheath expansion can be explained by electrons produced by the NERH at the sides of the racetrack region, which then transit through the EMCR and become confined by the magnetic field there (such as the ``blue'' particle in Fig.\ref{Fig8a}a). The $P_{p,z}$ component can be attributed to the interaction of the EMCR electron population with the time-modulated magnetized sheath underlying the MSH \cite{lieberman_1991}. One can see that the corresponding electron heating (cooling) rate is comparable to those exhibited by $P_{p,\theta}$ and $P_{i,\theta}$. However, if one looks at the corresponding period-averaged contributions of different components in the mechanical energy balance (see Fig.\ref{Fig10c}), one can see that the average contribution of the MSH is small compared to the terms related to the Hall heating. A more detailed discussion of $P_{p,\theta}$ and $P_{i,\theta}$ should respect the fact that these terms are parts of the gyroforce resulting from the divergence of the gyroviscous tensor. However, the expressions for the gyroforce are notoriously complicated \cite{smolyakov_1998,ramos_2005} and such a discussion would go beyond the scope of the present paper.

\begin{figure}[h]
\centering
\includegraphics[width=12cm]{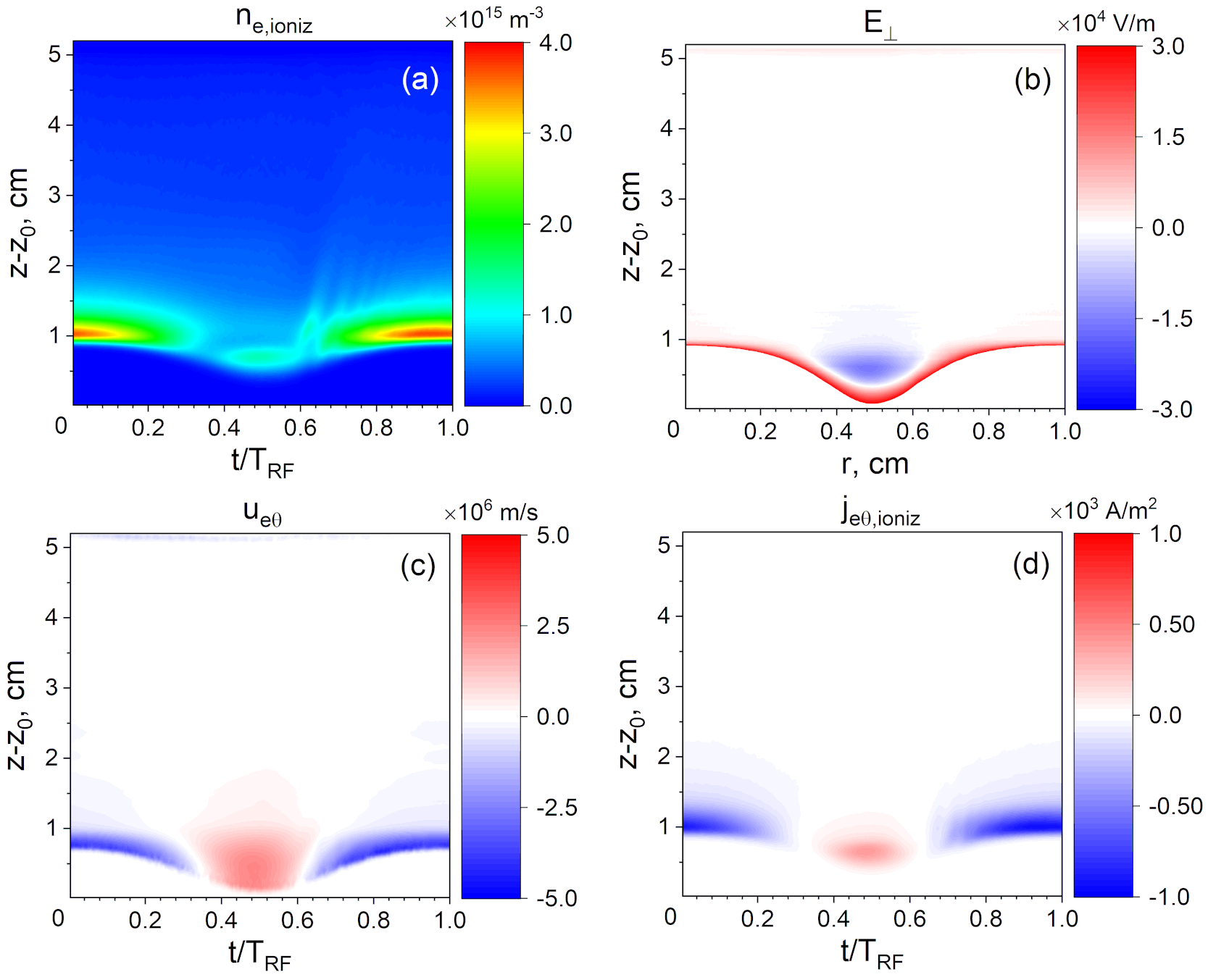}
\caption{Temporal evolution of the energetic (with energy above the ionization threshold of $15.8$ eV) electrons' axial density profile (a), axial profile of the transverse electric field (b), axial profile of the average azimuthal velocity component (c), and the axial profile of the azimuthal energetic electrons' current density (d) sampled at the radial racetrack location $r=2.7$ cm.}
\label{Fig10a}
\end{figure}

There are two major phases of the discharge dynamics, where an intensive energy exchange between electrons and the electric field takes place - during the sheath collapse, where a strong reversed electric field is generated in the transverse current channel part above the racetrack to facilitate electron transport in the corresponding channel similar to \cite{eremin_2021,wang_2020},
ultimately leading to the powered electrode (see Fig.\ref{Fig10a}b), and during the sheath expansion/retreat. 
To see that such electrons experience strong ${\bf E \times B}$ drift we investigated the average azimuthal velocity, which can be observed in Fig.\ref{Fig10a}c: the average velocity becomes as large as $5\times 10^6$ m/s, which corresponds to $71$ eV. However, the largest azimuthal velocities are attained in regions where electron densities are small.
Therefore, for a more accurate estimate of the energetic electron population generated due to the azimuthal Hall heating, one should take into account the electron density profile, which can be done by calculating the profile of azimuthal electron current density produced by the energetic electrons. The corresponding data is shown in Fig.\ref{Fig10a}d, which agrees well with the energetic electron population shown in Fig.\ref{Fig10a}a. This also agrees with the assumption that the dominant energetic electron population in the confinement region above the racetrack arises primarily due to the Hall heating/energization mechanism. 

\begin{figure}[h]
\centering
\includegraphics[width=16cm]{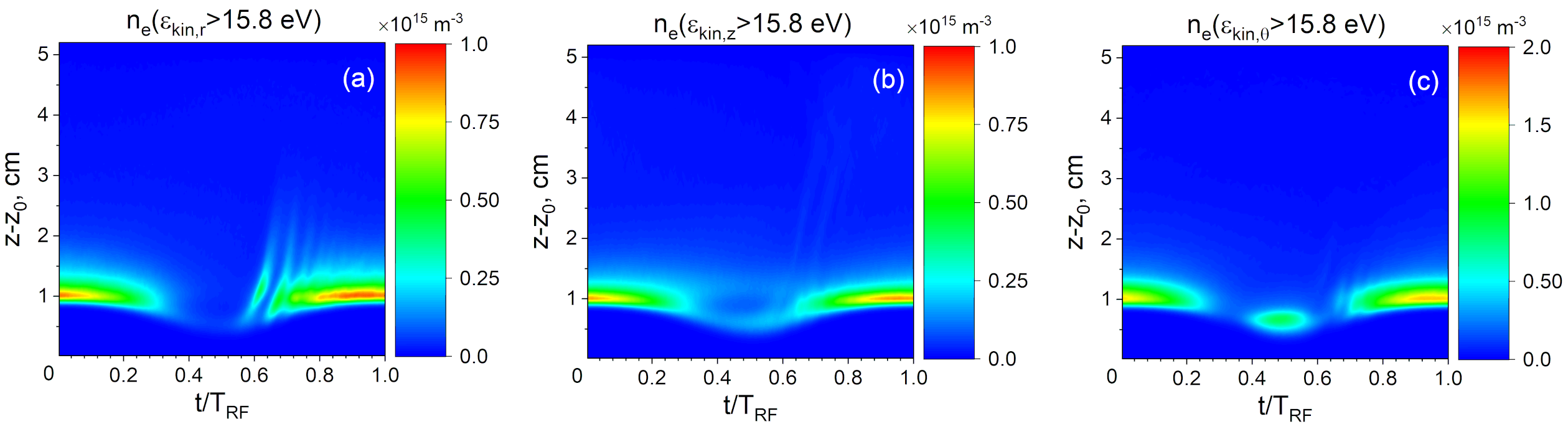}
\caption{Temporal evolution of the axial density profiles calculated for electrons with the kinetic energy associated with only one of the three velocity components being above the ionization threshold. The density profiles are sampled at the radial racetrack location $r=2.7$ cm.}
\label{Fig10b}
\end{figure}

One can also appreciate the Hall heating's significance by examining Fig.\ref{Fig10b}, which shows the temporal evolution of the axial density profiles calculated for electrons with the kinetic energy associated with only one of the three velocity components being above the ionization threshold.
The energetic electron densities calculated for the radial and axial velocity components (Fig.\ref{Fig10b}a and Fig.\ref{Fig10b}b) are comparable and show similar dynamics. This indicates that the electron velocity distribution in the $(v_r,v_z)$ plane is approximately isotropic. Since electrons which are sampled at this radial location come mostly from the energetic electrons trapped in the EMCR, this is to be expected. Note, however, that Fig.\ref{Fig10b}a shows strong beams of energetic electrons during the start of the sheath expansion. This can be linked to the energetic electrons produced by NERH, which propagate mostly along the magnetic field lines, which point in the radial direction above the racetrack location. Some of those electrons are too energetic to remain confined by the magnetic field and leave the EMCR, but the rest becomes retained and replenishes the electron 
population there (see the blue particle's trajectory in Fig.\ref{Fig8a}a).
In contrast, Fig.\ref{Fig10b}c exhibits density of energetic electrons much larger (approximately by a factor of two) than that seen in Fig.\ref{Fig10b}a and Fig.\ref{Fig10b}b. In addition, it is the only plot that demonstrates significant production of energetic electrons during the sheath collapse. This can only be explained by strong Hall heating associated with the transverse electric field (see Fig.\ref{Fig10a}b and the associated discussion). 

\subsection{Electron energization and excitation rate dynamics \label{ss3d}}

\begin{figure}[h]
\centering
\includegraphics[width=17cm]{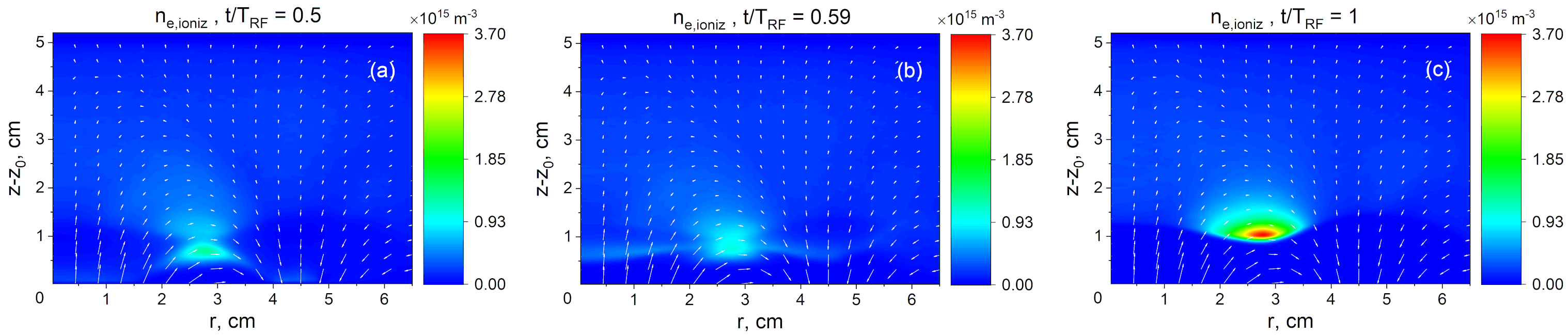}
\caption{Snapshots of the density of energetic electrons sampled at moments during the sheath collapse (a), close to the begin of the sheath expansion (b), and at the maximum of the sheath expansion (c).}
\label{Fig11}
\end{figure}

To see what energization mechanism is the most efficient in the production of energetic electrons above the ionization threshold, Fig.\ref{Fig11} shows the energetic electron density at the sheath collapse (a), during the start of the sheath expansion (b), and during the advanced sheath expansion phase (c). 
One can observe that the longitudinal electric field reversal heating occurring close to the target during
the sheath collapse produces only a small fraction of the energetic electrons. It is also due to the fact that a large fraction of energetic electrons there is lost to the powered electrode to maintain the overall charge balance. One can also observe
that there is a group of energetic electrons flowing from the EMCR along the boundary of the magnetized sheath in the direction of the powered electrode.
Two dominant populations of energetic electrons can be observed, which are confined by the magnetic field-related mirror effect (note that the electrostatic sheath is collapsed at that moment), both in the racetrack region. They experience strong heating by the Hall mechanism caused by the reversed transverse electric field generated in the transverse current channel above the racetrack.  

There are energetic electron beams produced by the moving sheath edge during the sheath expansion's starting phase (see Fig\ref{Fig11}b). The expanding sheath rapidly oscillates and pushes electrons into the bulk along the magnetic field lines. Some electrons are energized by the ambipolar field at the pre-sheath edge. But the dominant energetic electron population is again in the EMCR above the racetrack and experiences the Hall heating generated by the transverse electric field. 

\begin{figure}[h]
\centering
\includegraphics[width=6.5cm]{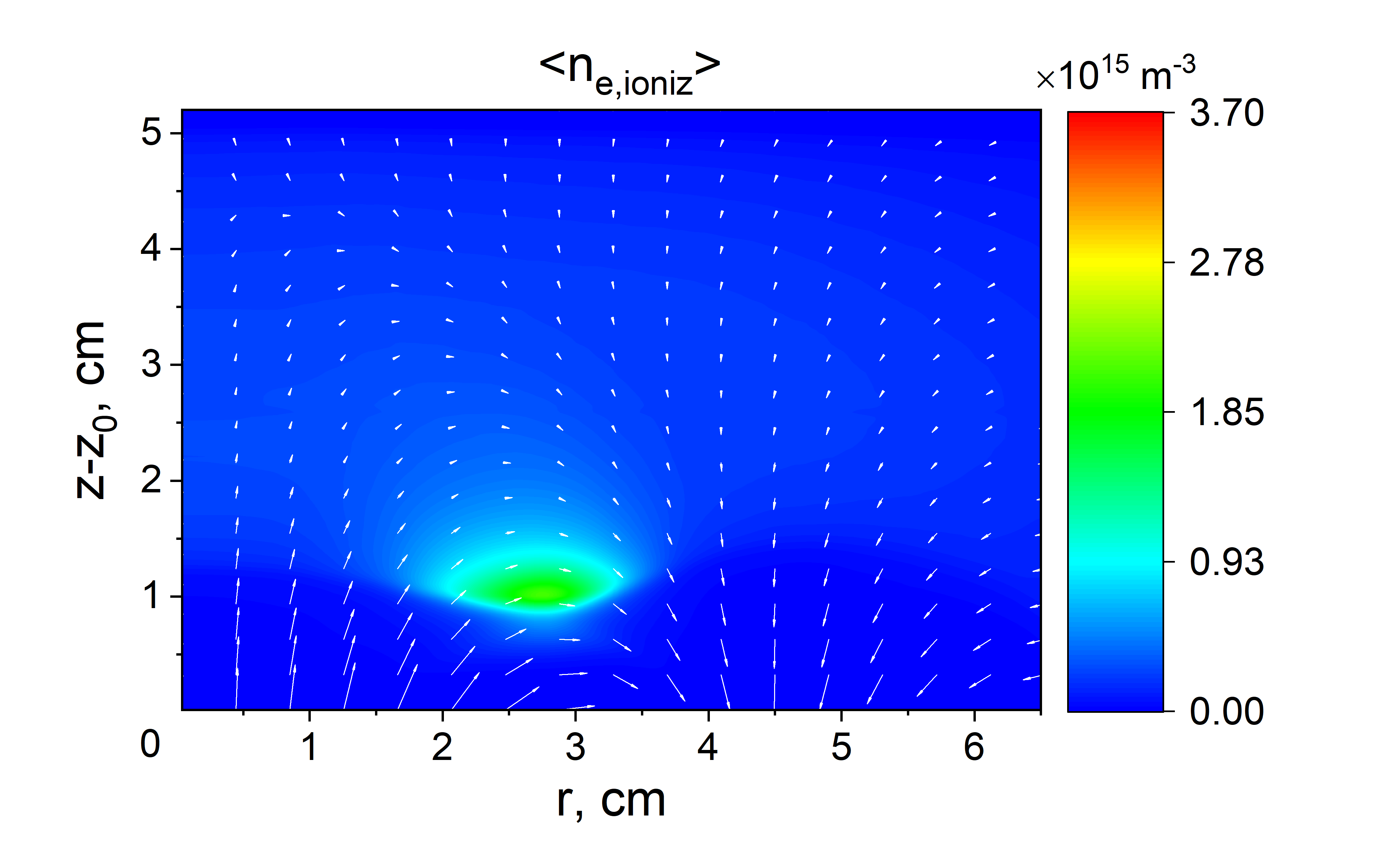}
\caption{Time-averaged density of energetic electrons.}
\label{Fig5}
\end{figure}

Finally, as the sheath is well expanded (see Fig\ref{Fig11}c), the Hall heating in the EMCR produces a very large population of energetic electrons, which makes the dominant contribution to the time-averaged energetic electron density profile (see Fig.\ref{Fig5}).

The dominance of the Hall heating in the overall electron energization over the rf period can be attributed to two reasons. Firstly, the Hall heating is very efficient in accelerating particles to very large energies by generating a strong azimuthal ${\bf E \times B}$ drift. Secondly, the Hall heating heats predominantly electrons, which are well confined by the magnetic field and remain in the reactor for a long time. Although the longitudinal NERH heating mechanism also produces highly energetic electrons with comparable energies, most of them are quickly lost to the electrodes (such as the ``magenta'' particle in Fig.\ref{Fig8a}) and do not significantly contribute to the average energetic electron population. 

%


\section{Conclusions} \label{sec5}

In a series of companion papers (see also \cite{eremin_2021} and \cite{berger_2021}) we attempt to identify
various mechanisms in rfMS discharges operated at a low pressure with a conducting target, which lead to the electron heating, i.e., overall energy gain from the electric field, and/or the electron energization, i.e., energy transfer from the electric field to the electrons in the inelastic range.
To this end, in this paper we conduct a 2D PIC simulation of a typical argon rfMS discharge in an $(r,z)$ axisymmetric geometry of a planar magnetron. To make a distinction with the physics of dcMS 
discharges, we deliberately excluded one of the main electron energization mechanisms there, which is related to the acceleration of ion-induced secondary electrons in the sheaths. We demonstrate that the electron energization in rfMS discharges involves several possible mechanisms, which efficiently channel the electric field energy into energetic electrons. The patterns obtained in the numerical simulations of this work seem to be in good agreement with experimental observations made in the third companion paper \cite{berger_2021}. 

Since the power transfer from the electric field to electrons is proportional to the electron current density along the electric field, the power transfer dynamics
depends on details of the electron transport. The latter is not trivial, since close to the target electrons are well magnetized and tend to follow the magnetic field lines, but if their energy is sufficiently large, electrons moving in the considered non-uniform magnetic field can experience complicated orbits. In particular, such energetic electrons can significantly deviate from the magnetic field lines and penetrate the area close to the grounded electrode, where they can be considered unmagnetized. Following the methodology applicable to strongly magnetized plasmas, we find it instrumental to separate the electron dynamics along (longitudinal) and orthogonal (transverse) to the magnetic field lines. For the typical topology of the planar magnetron's magnetic field we show that there could be various electron current channels depending on the character of the electron motion in the region of their magnetization (close to the target). The transverse electron current channel consists of a section with predominantly transverse electron transport, which goes through the electron confinement region located above the racetrack, and a section along the magnetic field lines guiding electrons to the powered electrode. In contrast to this, the two longitudinal electron current channels located on both sides of the racetrack region guide electrons predominantly along the magnetic field lines. We find that during the sheath collapse all three electron current channels generate electron flow to the powered electrode in order to compensate the positive charge supplied by the flow of unmagnetized ions. To enhance the electron flow along the magnetic field lines close to the target in all channels, a reversed electric field is produced to overcome the mirror force. During the sheath expansion the electron current in the transverse channel flows smoothly through its transverse section, whereas the sheath edge in the longitudinal current channels exhibits rapid PSR-like oscillations. Such a disparity in the electron dynamics can be explained by different roles of the magnetic field in the channels. In particular, the magnetic field's impact in the longitudinal channels is small, so that the corresponding dynamics resembles that in unmagnetized CCRF discharges.  

Depending on the electron current channel, the electron heating/energization mechanisms can accordingly be split into the longitudinal and transverse mechanisms. The reversed electric field generated along the magnetic field lines during the sheath collapse to counteract the mirror force leads to an increase in the longitudinal part of the electron kinetic energy, so that enough electrons can reach the powered electrode. This electric field also heats other electrons in the corresponding regions close to the target, increasing their longitudinal energy, albeit not sufficiently to overcome the mirror force and arrive at the powered electrode. We suggest to call this new electron heating mechanism the ``mirror-effect heating'' (MEH). However, the dominant electron heating during this phase is generated in the transverse part of the transverse electron current channel. There, a reversed electric field generated in the regions above the racetrack, where electrons are well confined due to the mirror effect, leads to a strong Hall heating (see a brief description in the Introduction and the companion paper \cite{eremin_2021}). During the sheath expansion the expanding sheath edge in the longitudinal electron current channels, which simultaneously undergoes rapid oscillations, accelerates electrons along the magnetic field lines to large energies. In this case it is similar to what happens in the NERH electron heating mechanism in unmagnetized CCRF discharges (note that the transverse electron current channel does not exhibit such sheath edge oscillations and has no related NERH electron heating). A large fraction of such energetic electrons have trajectories aligned with the magnetic field lines or detaching from them, leading the energetic electrons to the oppositely located grounded electrode, where they leave the discharge. Such electrons are too energetic to be confined by the effective potential consisting of the electric and the magnetic parts, the latter stemming from the conservation of the azimuthal canonical momentum \cite{sheridan_1989}. However, we show that some of the energetic electrons become trapped in the magnetic field and can undergo several energizing interactions with the expanding sheath in the longitudinal current channels on both sides of the racetrack region. In this new mechanism, which we propose to call ``magnetized bounce heating'' (MBH) electrons bounce from one region to another along the guiding arc-like magnetic field. Some of the additional longitudinal heating seems to take place due to the ambipolar electric field similar to unmagnetized CCRF discharges. The dominant electron energization mechanism during the sheath expansion turns out to be the Hall heating caused by the generation of a strong ${\bf E \times B}$ drift in the azimuthal direction by the transverse electric field in the transverse electron current channel. 
In general, certain interaction between the energetic electrons produced in the longitudinal and the transverse channel is possible. On the one hand, the magnetic field can trap some of the electrons energized by the NERH. On the other hand, some of electrons with relatively small pitch angle can flow close to the powered electrode from the EMCR during the sheath collapse and increase their longitudinal energy due to the NERH, flowing back into the discharge along a longitudinal channel. 



\section*{Acknowledgments}

The authors gratefully acknowledge support by DFG (German Research Foundation) within the framework
of the Sonderforschungsbereich SFB-TR 87 and the project
``Plasmabasierte Prozessführung von reaktiven Sputterprozessen'' (No. 417888799).  

\appendix
\section{Magnetic field approximation} \label{Appendix_a}

The magnetic flux density components are approximated by the vacuum-series solution from  experimental data,
\begin{equation}
\begin{array}{lll}
&& B_r(r,z) = \sum\limits_{i=1}^n A_i J_1(\lambda_i r) \exp(-\lambda_i |z-z_0|) \\ 
&& B_z(r,z) = \sum\limits_{i=1}^n A_i J_0(\lambda_i r) \exp(-\lambda_i |z-z_0|)
\end{array} \label{eqA_1}
\end{equation}
with $n=4$ and
\begin{equation}
\begin{array}{lll}
A_1 = -3.2765 \, {\rm mT} &,& \lambda_1 = 35.3677 \, {\rm m}^{-1} \\
A_2 = 25.0918 \, {\rm mT} &,& \lambda_2 = 81.1765 \, {\rm m}^{-1} \\
A_3 = -3.1731 \, {\rm mT} &,& \lambda_3 = 127.2603 \, {\rm m}^{-1} \\
A_4 = -7.2414 \, {\rm mT} &,& \lambda_4 = 173.4044 \, {\rm m}^{-1} 
\end{array}
\end{equation}
for the ``$7$ mT'' magnetic field configuration considered in the present work (the reference value is related to the approximate magnetic flux density measured in the 
experiment $8$ mm above the racetrack radial location).

\section{Mechanical energy balance diagnostics} \label{Appendix_b}

Mechanical energy balance diagnostics \cite{surendra_1993} can be helpful for the identification of different mechanisms contributing to the electron heating (e.g., \cite{zheng_2019,lafleur_2014,schulze_2018}). Here we generalize its derivation made in \cite{eremin_2021} for cylindrical coordinates in 1D radial geometry to the 2D radial-axial geometry.  

The Boltzmann equation for the chosen geometry can be obtained by using Eqs.(\ref{eq2_1}) for the characteristics, which yields
\begin{equation}
\begin{array}{lll}
\frac{\partial f_e}{\partial t} + v_r\frac{\partial f_e}{\partial r}
 + v_z\frac{\partial f_e}{\partial z} \\
-\left[\frac{e}{m_e}(E_r + v_\theta B_z) -\frac{v_\theta^2}{r}\right]\frac{\partial f_e}{\partial v_r} 
- \left[\frac{e}{m_e}(v_z B_r - v_r B_z)  + \frac{v_\theta v_r}{r}\right]\frac{\partial f_e}{\partial v_\theta} \\
-\frac{e}{m_e}(E_z - v_\theta B_r)\frac{\partial f_e}{\partial v_z}
= C(f_e),
\end{array}
\label{eqB_1}
\end{equation}
where $C(f_e)$ is the collision operator for electrons describing the rate of the electron distribution 
function's change in phase space due to collisions.  Multiplying this equation by $v_r$ and integrating over velocity space yields the radial momentum balance equation
\begin{equation}
\begin{array}{lll}
m_e\frac{\partial}{\partial t} (n_eu_{e,r}) + \frac{m_e}{r}\left(\frac{\partial}{\partial r}(rn_eu_{e,r}^2) - n_e u_{e,\theta}^2\right) +m_e\frac{\partial}{\partial z}(n_eu_{e,r}u_{e,z}) \\
+\frac{1}{r}\frac{\partial}{\partial r}(r p_{e,rr}) - \frac{p_{e,\theta\theta}}{r} + \frac{\partial}{\partial z}p_{zr} \\
+ e n_e E_r + en_eu_{e,\theta}B_z
= m_e \int C(f_e) v_r d{\bf v}, 
\label{eqB_2}
\end{array}
\end{equation}
with the electron density $n_e  = \int f_e d{\bf v}$, average velocity components $u_{e,i} = n_e^{-1}\int f_e v_i d{\bf v}$, and 
the pressure tensor components $p_{e,i j} = m_e\int f_e (v_i-u_{e,i})(v_j-u_{e,j}) d{\bf v}$. One can use this equation to express the electric field's radial component, which provides
\begin{equation}
\begin{array}{lll}
E_r = \\
-\frac{1}{n_e}\left[ m_e\frac{\partial}{\partial t} (n_eu_{e,r}) + \frac{m_e}{r}\left(\frac{\partial}{\partial r}(rn_eu_{e,r}^2) + n_e u_{e,\theta}^2\right) 
+m_e\frac{\partial}{\partial z}(n_eu_{e,r}u_{e,z})
\right. \\ 
\left.
+\frac{1}{r}\frac{\partial}{\partial r}(r p_{e,rr}) - \frac{p_{e,\theta\theta}}{r} + \frac{\partial}{\partial z}p_{zr} 
+ en_eu_{e,\theta}B_z
- m_e \int C(f_e) v_r d{\bf v} \right]. 
\label{eqB_3}
\end{array}
\end{equation}
A similar procedure conducted in the $z$ direction leads to
\begin{equation}
\begin{array}{lll}
E_z = \\
-\frac{1}{n_e}\left[ m_e\frac{\partial}{\partial t} (n_eu_{e,z}) + \frac{m_e}{r}\frac{\partial}{\partial r}(rn_eu_{e,r}^2) 
+m_e\frac{\partial}{\partial z}(n_eu_{e,z}^2)
\right. \\ 
\left.
+\frac{1}{r}\frac{\partial}{\partial r}(r p_{e,rz})  + \frac{\partial}{\partial z}p_{zz} 
- en_eu_{e,\theta}B_r
- m_e \int C(f_e) v_z d{\bf v} \right]. 
\label{eqB_4}
\end{array}
\end{equation}
The azimuthal momentum balance equation reads
\begin{equation}
\begin{array}{lll}
m_e\frac{\partial}{\partial t} (n_eu_{e,\theta}) + \frac{m_e}{r}\left(\frac{\partial}{\partial r}(rn_eu_{e,r}u_{e,\theta}) + n_e u_{e,r}u_{e,\theta}\right) +m_e\frac{\partial}{\partial z}(n_eu_{e,z}u_{e,\theta}) \\
+\frac{1}{r}\frac{\partial}{\partial r}(r p_{e,r\theta}) + \frac{p_{e,r\theta}}{r} + \frac{\partial}{\partial z}p_{z\theta} \\
+ en_e(u_{e,z}B_r - u_{e,r}B_z)
= m_e \int C(f_e) v_\theta d{\bf v}. 
\label{eqB_5}
\end{array}
\end{equation}
The power density absorbed by electrons from the electric field is equal to \begin{equation}
P_{e,tot} = -en(u_{e,r}E_r + u_{e,z}E_z). \label{eqB_6} 
\end{equation}
It is evident that if one uses Eqs.(\ref{eqB_3}) and (\ref{eqB_4}) to express it via the momentum balance terms, it will contain contributions from the magnetic field components (sometimes called Hall terms, see, e.g., \cite{zheng_2021}). However, the magnetic field does not perform work on particles, since the corresponding Lorentz force is always orthogonal to the particle velocities. Therefore, the final expression for the power absorbed by electrons should not contain any Hall terms. Indeed, one can see that the corresponding terms can be completely excluded with help of the azimuthal balance equation (\ref{eqB_5}) multiplied with $u_{e,\theta}$. Performing this procedure yields the final expression for $P_{e,tot}$,
\begin{equation}
P_{e,tot} = P_{i,r} + P_{i,\theta} + P_{i,z} 
+ P_{p,r} + P_{p,\theta} + P_{p,z}
+ P_{c,r} + P_{c,\theta} + P_{c,z}
\label{eqB_7}
\end{equation}
with $P_i$ being contributions from the "inertial", $P_p$ the "pressure", and $P_c$ the collisional mechanisms, and
$$
P_{i,r} = 
m_eu_{e,r}\frac{\partial}{\partial t} (n_eu_{e,r}) + \frac{m_eu_{e,r}}{r}\left(\frac{\partial}{\partial r}(rn_eu_{e,r}^2) - n_e u_{e,\theta}^2\right) 
+m_eu_{e,r}\frac{\partial}{\partial z}(n_eu_{e,r}u_{e,z}),
$$
$$
P_{i,\theta} = m_eu_{e,\theta}\frac{\partial}{\partial t} (n_eu_{e,\theta}) + \frac{m_eu_{e,\theta}}{r}\left(\frac{\partial}{\partial r}(rn_eu_{e,r}u_{e,\theta}) + n_e u_{e,r}u_{e,\theta}\right) +m_eu_{e,\theta}\frac{\partial}{\partial z}(n_eu_{e,z}u_{e,\theta}),
$$
$$
P_{i,z} = m_eu_{e,z}\frac{\partial}{\partial t} (n_eu_{e,z}) + \frac{m_eu_{e,z}}{r}\frac{\partial}{\partial r}(rn_eu_{e,r}^2) 
+m_eu_{e,z}\frac{\partial}{\partial z}(n_eu_{e,z}^2),
$$
$$
P_{p,r} = 
\frac{u_{e,r}}{r}\frac{\partial}{\partial r}(r p_{e,rr}) - u_{e,r}\frac{p_{e,\theta\theta}}{r} + u_{e,r}\frac{\partial}{\partial z}p_{zr}
,
$$
$$
P_{p,\theta} =
\frac{u_{e,\theta}}{r}\frac{\partial}{\partial r}(r p_{e,r\theta}) + u_{e,\theta}\frac{p_{e,r\theta}}{r} + u_{e,\theta}\frac{\partial}{\partial z}p_{z\theta},
$$
$$
P_{p,z} =
\frac{u_{e,z}}{r}\frac{\partial}{\partial r}(r p_{e,rz})  + u_{e,z}\frac{\partial}{\partial z}p_{zz},
$$
$$
P_{c,r} = m_e u_{e,r} \int C(f_e) v_r d{\bf v},
$$
$$
P_{c,\theta} = m_e u_{e,\theta} \int C(f_e) v_\theta d{\bf v},
$$
and
$$
P_{c,z} = m_e u_{e,z} \int C(f_e) v_z d{\bf v}.
$$

It is important to note that although the collisional contributions are virtually always attributed to the Ohmic heating mechanism (e.g.,\cite{zheng_2019}), one should be careful in their interpretations. As we demonstrate in \cite{eremin_2021}, in case of the magnetized plasmas
the dominating electron heating mechanism, which we suggest to call the Hall heating, can be essentially collisionless and is caused by the generation of a strong ${\bf E}\times{\bf B}$ drift (in the azimuthal direction for the geometry considered here) producing a significant fraction of energetic electrons in the inelastic energy range. The collisional signature $P_{c,\theta}$ in the corresponding direction is mainly due to the conversion of the energy associated with azimuthal drift of energetic electrons into the thermal energy through elastic collisions and due to the inelastic collisions with large energy losses.

%
%

\section*{References}
\bibliographystyle{ieeetr}

\bibliography{references}

\end{document}